\documentclass[10pt,manuscript,screen]{acmart}
\usepackage{bm}
\usepackage{algorithmic}
\usepackage{graphicx}
\usepackage{textcomp}
\usepackage{xcolor}
\usepackage{fancyhdr}
\usepackage{multirow}
\usepackage{dsfont}
\usepackage{subcaption}

\usepackage{pifont}% http://ctan.org/pkg/pifont

\usepackage{tablefootnote}
\usepackage{tabularx}
\usepackage{footnote}
\makesavenoteenv{tabular}

\newcommand{\cmark}{\ding{51}}%
\newcommand{\xmark}{\ding{55}}%
\AtBeginDocument{%
  \providecommand\BibTeX{{%
    \normalfont B\kern-0.5em{\scshape i\kern-0.25em b}\kern-0.8em\TeX}}}

\begin{document}
\title{QUIDAM: A Framework for \underline{Qu}ant\underline{i}zation-Aware \underline{D}NN \underline{A}ccelerator and \underline{M}odel Co-Exploration}
\author{Ahmet Inci}
\email{ainci@andrew.cmu.edu}
\affiliation{%
  \institution{Carnegie Mellon University}
  \streetaddress{5000 Forbes Avenue}
  \city{Pittsburgh}
  \state{PA}
  \country{USA}
  \postcode{15213}
}

\author{Siri Garudanagiri Virupaksha}
\affiliation{%
  \institution{Carnegie Mellon University}
  \streetaddress{5000 Forbes Avenue}
  \city{Pittsburgh}
  \state{PA}
  \country{USA}
  \postcode{15213}
  }
% \email{sgarudan@andrew.cmu.edu}

\author{Aman Jain}
\affiliation{%
  \institution{Carnegie Mellon University}
  \streetaddress{5000 Forbes Avenue}
  \city{Pittsburgh}
  \state{PA}
  \country{USA}
  \postcode{15213}
}

\author{Ting-Wu Chin}
\affiliation{%
  \institution{Carnegie Mellon University}
  \streetaddress{5000 Forbes Avenue}
  \city{Pittsburgh}
  \state{PA}
  \country{USA}
}

\author{Venkata Vivek Thallam}
\affiliation{%
  \institution{Carnegie Mellon University}
  \streetaddress{5000 Forbes Avenue}
  \city{Pittsburgh}
  \state{PA}
  \country{USA}}

\author{Ruizhou Ding}
\affiliation{%
  \institution{Carnegie Mellon University}
  \streetaddress{5000 Forbes Avenue}
  \city{Pittsburgh}
  \state{PA}
  \country{USA}}

\author{Diana Marculescu}
\email{dianam@utexas.edu}
\affiliation{%
  \institution{Carnegie Mellon University,}
  \streetaddress{5000 Forbes Avenue}
  \city{Pittsburgh}
  \state{PA}
   \postcode{15213}
   \institution{The University of Texas at Austin}
  \streetaddress{110 Inner Campus Drive}
  \city{Austin}
  \state{TX}
  \country{USA}
   \postcode{78705}}

\renewcommand{\shortauthors}{Inci et al.}

\begin{abstract}
As the machine learning and systems communities strive to achieve higher energy-efficiency through custom deep neural network (DNN) accelerators, varied precision or quantization levels, and model compression techniques, there is a need for design space exploration frameworks that incorporate quantization-aware processing elements into the accelerator design space while having accurate and fast power, performance, and area models. In this work, we present \textit{QUIDAM}, a highly parameterized quantization-aware DNN accelerator and model co-exploration framework. Our framework can facilitate future research on design space exploration of DNN accelerators for various design choices such as bit precision, processing element type, scratchpad sizes of processing elements, global buffer size, number of total processing elements, and DNN configurations. Our results show that different bit precisions and processing element types lead to significant differences in terms of performance per area and energy. Specifically, our framework identifies a wide range of design points where performance per area and energy varies more than $5 \times$ and $35 \times$, respectively. With the proposed framework, we show that lightweight processing elements achieve on par accuracy results and up to $5.7 \times$ more performance per area and energy improvement when compared to the best INT16 based implementation. Finally, due to the efficiency of the pre-characterized power, performance, and area models, QUIDAM can speed up the design exploration process by 3-4 orders of magnitude as it removes the need for expensive synthesis and characterization of each design.

%\textit{QUIDAM} can reduce DNN accelerator and model co-exploration time by orders of magnitude by using our quantization-aware power, performance, and area models which shows the efficacy of our proposed framework.

\end{abstract}



% \keywords{datasets, neural networks, gaze detection, text tagging}

\maketitle

\section{Introduction}\label{sec:quidam_intro}
Deep neural networks (DNNs) have achieved remarkable accomplishments across various applications ranging from image recognition \cite{EfficientNet}, object detection \cite{EfficientDet}, to natural language processing \cite{Devlin2019BERTPO}. However, the increasing model size and computational cost of these models has become a challenging task for on-device machine learning (ML) endeavours due to the stringent performance per area and energy constraints of the edge devices. To this end, while machine learning practitioners focus on model compression techniques \cite{han2015deep_compression,ruizhou2018lightnn,Chin2020CVPR}, computer architects investigate hardware architectures to overcome the energy-efficiency problem and improve the overall system performance \cite{eyeriss,aly2015next,han2016eie,chen2017dataflow,Shao2019SimbaSD,inci2018asbd,DeepNVM,inci2021tcad,inci2020architectural,inci2022qappa,inci2021cross,inci2021qadam,inci2022nvmbookchapter}.

As computing community hits the limits on consistent performance scaling for traditional architectures, there has been a rising interest on enabling on-device machine learning applications through custom domain-specific DNN accelerators. These domain-specific system-on-chip architectures are specifically designed to exploit the application characteristics. As we deeply care about performance per area and energy-efficiency from a hardware point of view, tailored DNN accelerators have shown significant improvements when compared to general-purpose CPUs and GPUs \cite{eyeriss,tpu,tpuv4,samsung_npu,Parashar2017SCNNAA,tetris}. To better understand the trade-offs of various architectural design choices and DNN workloads for domain-specific hardware architectures, there is a need for a design space co-exploration framework that can rapidly iterate over various designs and generate power, performance, and area (PPA) results for various DNN models. To this end, in this work we present \textit{QUIDAM}, a framework for quantization-aware DNN accelerator and model co-exploration.

This work makes the following contributions: 
\begin{itemize}
    \item We present \textit{QUIDAM}, a quantization-aware power, performance, and area modeling framework for DNN accelerators. Our framework can enable future research on DNN accelerator and model co-exploration for various design choices such as bit precision, processing element types, scratchpad sizes of processing elements, global buffer size, device bandwidth, number of total processing elements, and DNN architectures. 
    \item  Our framework provides power, performance, and area results not just for a single hardware design point but for a range of different hardware designs as opposed to prior art \cite{qi17paleo,cai2017neuralpower,eyeriss,eyeriss_ssc,maeri}. Thus, it can be used to jointly analyze trade-offs of various architectural design choices and DNN workloads and perform multi-objective optimization to achieve a better trade-off front between accuracy and hardware-efficiency metrics such as performance per area and energy. 
\end{itemize}

The rest of the paper is organized as follows. In Section \ref{sec:quidam_relatedwork}, we present a literature review on power and runtime models for CNNs and design space exploration frameworks for hardware accelerators. 
In Section \ref{sec:quidam_methodology}, we describe the architectural details of the \textit{QUIDAM} framework and the details of our methodology for power, performance, and area modeling of DNN accelerators. In Section \ref{sec:quidam_results}, we show experimental results demonstrating the efficacy of \textit{QUIDAM}'s power, performance, and area models and the efficacy of lightweight processing elements to conventional designs in terms of performance per area and energy through a suite of case studies. 
Finally, Section \ref{sec:quidam_conclusion} concludes the paper by summarizing the results.

\section{Related Work}\label{sec:quidam_relatedwork}

\begin{figure}[t]
  \centering
 \includegraphics[width=0.8\textwidth]{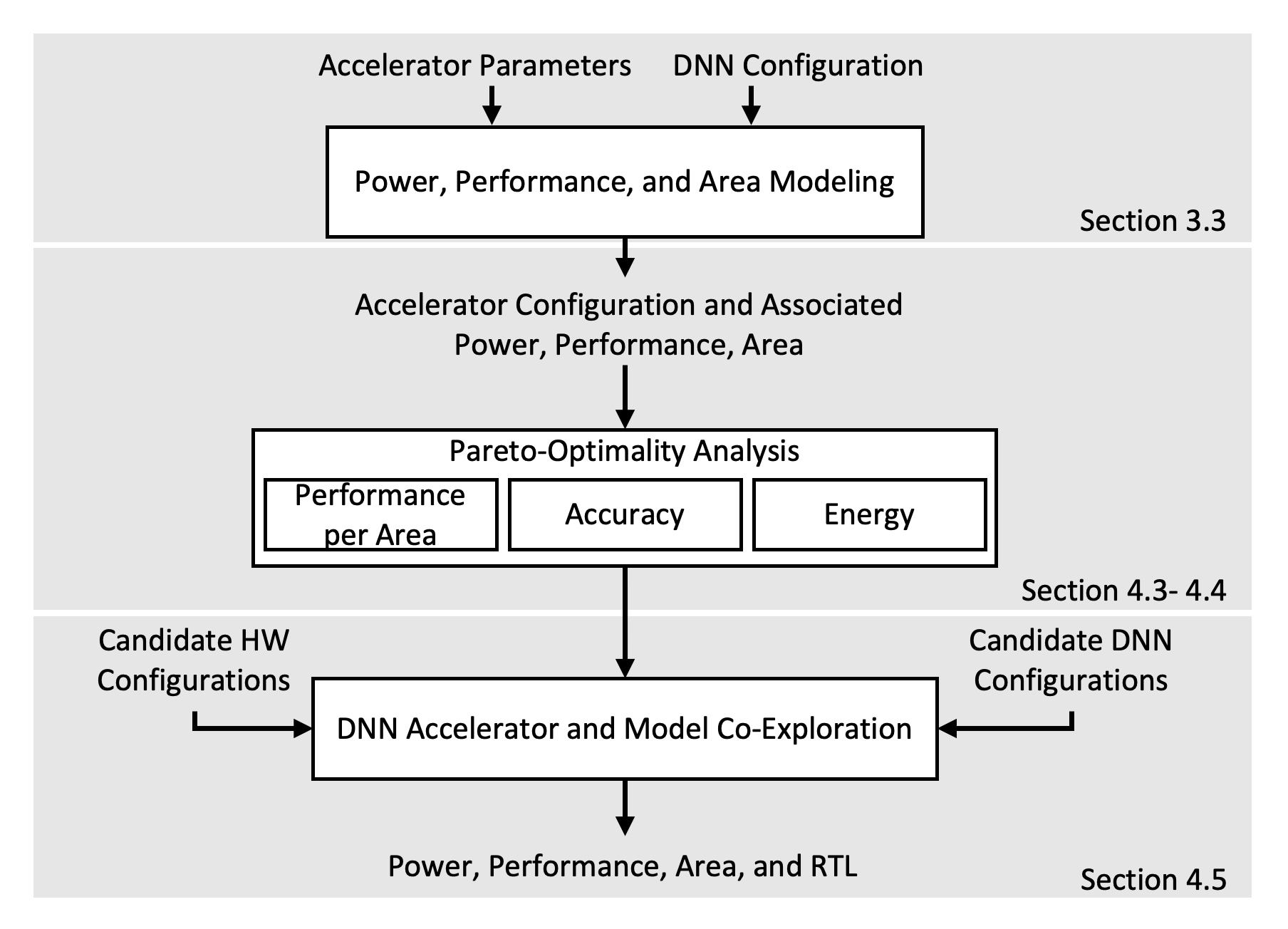}
 \caption{{Overview of the \textit{QUIDAM} framework.}}
 \label{fig:flow}
\end{figure}

Prior art has proposed runtime and energy models for DNN workloads \cite{cai2017neuralpower,qi17paleo,marculescu2018modelling}. However, these models have been implemented specifically for GPU platforms and thus they create an important limitation for a design space exploration of hardware architectures and potentially hardware and machine learning model co-design opportunities \cite{zhou2021rethinking,gupta2020acceleratoraware,codesign_vikas}.

\begin{table}[]
\centering
\caption{Comparison of DNN Accelerator Frameworks}
\label{table:comparison_table}
\resizebox{1\textwidth}{!}{%
\begin{tabular}{||c|c|c|c|c|c|c|c||}
\hline
\multirow{2}{*}{} &
  \multirow{2}{*}{\begin{tabular}[c]{@{}c@{}}Optimized for\\CNNs\end{tabular}} &
  \multirow{2}{*}{\begin{tabular}[c]{@{}c@{}}Fully-\\ Parameterized RTL\end{tabular}} &
  \multirow{2}{*}{\begin{tabular}[c]{@{}c@{}}Row-Stationary\\ Dataflow\end{tabular}} &
    \multirow{2}{*}{\begin{tabular}[c]{@{}c@{}}Quantization\\ Support\end{tabular}} &
  \multirow{2}{*}{\begin{tabular}[c]{@{}c@{}}Lightweight PE\\ (Shift-Add Based)\end{tabular}} &
  \multirow{2}{*}{\begin{tabular}[c]{@{}c@{}}Pre-Characterized \\ Model based DSE\end{tabular}} &
  \multirow{2}{*}{\begin{tabular}[c]{@{}c@{}}DNN Accelerator/\\ Model Co-Exploration\end{tabular}} \\
        &   &   &   &   &   &   &   \\ \hline \hline
Aladdin \cite{Shao2014AladdinAP} & \xmark & \xmark & \xmark & \xmark & \xmark & \cmark & \xmark \\ \hline
SCALE-Sim \cite{samajdar2018scale} & \cmark & \xmark & \xmark & \xmark & \xmark & \cmark & \xmark \\ \hline
Eyeriss \cite{eyeriss}      & \cmark & \xmark & \cmark & \xmark & \xmark & \xmark & \xmark \\ \hline
MAERI \cite{maeri}      & \cmark & \xmark & \cmark & \xmark & \xmark & \xmark & \xmark \\ \hline
MAESTRO \cite{maestro}      & \cmark & \xmark & \cmark & \xmark & \xmark & \cmark & \xmark \\ \hline
HAQ \cite{haq}      & \cmark & \xmark & \xmark & \cmark & \xmark & \xmark & \xmark \\ \hline
Accelergy \cite{accelergy}      & \cmark & \xmark & \cmark & \cmark$^{1}$ & \xmark & \xmark & \xmark \\ \hline
Timeloop \cite{timeloop}       & \cmark & \xmark & \cmark & \cmark$^{2}$ & \xmark & \cmark & \xmark \\ \hline
Gemmini \cite{gemmini}      & \cmark & \cmark & \xmark & \cmark$^{3}$ & \xmark & \xmark & \xmark \\ \hline
QUIDAM (Ours)       & \cmark & \cmark & \cmark & \cmark$^{4}$ & \cmark & \cmark & \cmark \\ \hline
\end{tabular}
}
\newline
\raggedright
\footnotesize{
\\ 1: Supports INT8, INT16, FP32 \\ 2: Supports INT8, INT16, FP32 \\ 3: Supports INT8, UINT32, FP32 \\ 4: Supports INT4, INT8, INT16, FP32}
\end{table}

On the other hand, the systems community has proposed several tools and simulation methodologies for DNN accelerator design. Table \ref{table:comparison_table} shows a comparison of existing hardware accelerator frameworks compared to \textit{QUIDAM} in terms of various design features in chronological order. 

For example,  Aladdin \cite{Shao2014AladdinAP} is a pre-RTL power, performance, and area estimation tool for arbitrary hardware accelerators. This simulator takes high-level language descriptions of algorithms as inputs similar to a high-level synthesis (HLS) methodology and generates dynamic data dependence graphs as an approximate representation of a hardware accelerator without generating RTL. While this approach can be useful for fast exploration of various algorithms, it has limitations in the optimization of the generated hardware accelerators for deep neural networks because it is not implemented in a domain-specific architecture manner. 

Similarly, SCALE-Sim \cite{samajdar2018scale} is a cycle accurate, systolic-array based DNN accelerator simulator. It has a python-based cycle accurate model that can generate results for hardware performance and utilization metrics. Although this tool can help rapid exploration of systolic-array based DNN accelerators for a given DNN layer, the built-in model is difficult to modify for a different hardware accelerator architecture and it lacks significant features such as quantization support, lightweight processing elements, and DNN accelerator and model co-exploration options. 

As mapping deep neural networks onto DNN accelerators plays an important role in energy-efficiency, Eyeriss \cite{eyeriss} analyzed various dataflow strategies in the existing DNN accelerator domain and came up with a novel approach called row-stationary dataflow which performs better than other dataflow strategies in terms of throughput and energy-efficiency \cite{eyeriss}. However, Eyeriss is only an instance in the vast design space of DNN accelerators and its implementation is not open-source to foster future research on DNN accelerator and model co-exploration. 

To this end, MAERI \cite{maeri} and MAESTRO \cite{maestro} have been proposed by the researchers to support spatial-array architectures and enrich the design space with various dataflows such as row-stationary dataflow which was originally proposed by Eyeriss \cite{eyeriss}. MAERI \cite{maeri} presents a new DNN accelerator architecture implemented with a set of modular and reconfigurable building blocks which can easily support various DNN mappings onto the accelerator by utilizing switches. On top of that, MAESTRO \cite{maestro} presents an analytical cost model to predict the hardware cost of dataflows to perform a design space exploration. These open-source frameworks assist researchers to perform a design space exploration on various dataflows. However, they are incapable of supporting various bit precision levels and efficient processing elements that utilize cheap shift logic instead of expensive multipliers.

To further improve the computational efficiency of hardware accelerators, researchers have investigated tuning the optimal bit precision level for each layer of a deep neural network. To this end, HAQ \cite{haq} proposed optimizing the bit precision of DNN accelerators by proposing an automated hardware-aware quantization framework that leverages reinforcement learning to automatically tune the quantization scheme for a given hardware architecture by integrating the power and performance feedback signals coming from the hardware architecture in the design loop. 

As HAQ reveals that the optimal bit precision levels on different hardware architectures and resource constraints are significantly different, recently Accelergy \cite{accelergy} and Timeloop \cite{timeloop} have been proposed to complement the design space exploration process by supporting different bit precision levels with an architecture-level energy estimation methodology \cite{accelergy} and a design space exploration framework \cite{timeloop} for DNN accelerators. 

Accelergy \cite{accelergy} presents an architecture-level energy estimation methodology for hardware accelerators that takes in an architecture description and hardware activity statistics such as action counts which are based on a given workload that is needed to be generated by a separate performance model. Although Accelergy \cite{accelergy} proposes a fast energy estimation methodology for DNN accelerators, it is not capable of performing a design space exploration for hardware architectures by itself. To this end, Timeloop \cite{timeloop} presents a framework for evaluating and performing a design space exploration of DNN accelerators. Timeloop \cite{timeloop} proposes a generic template to describe DNN hardware accelerators and characterize deep learning workloads and provides a practical design space exploration tool to understand the trade-offs in designing DNN accelerators across different workloads.

Although these tools perform preliminary analysis on the design space for DNN accelerators in different aspects, they do not incorporate specialized quantization-aware lightweight processing elements and they do not generate or share a highly-parameterized RTL implementation of the chosen design based on the input hardware configuration which is a significant impediment for enabling deployment of DNNs onto edge devices, as the actual deployment of the hardware design takes a significant amount of engineering effort.

Finally, Gemmini \cite{gemmini} has been proposed as an open-source co-processor/accelerator generator framework that leverages a flexible architectural template to represent different accelerator architectures. Moreover, Gemmini \cite{gemmini} provides a parameterized RTL implementation to enable hardware architects to gain more control onto the design and potentially gain subtle insights on how different architectural decisions affect overall performance of the system. To enable system researchers to investigate and run software stacks on the generated hardware accelerator, Gemmini \cite{gemmini} also supports a full system-on-chip environment. From the hardware implementation point of view, the internal DNN accelerator implementation is a systolic-array based architecture similar to SCALE-Sim \cite{samajdar2018scale} which is a relatively more simplistic architecture when compared to spatial-array based dataflow architectures \cite{eyeriss}. Therefore, it does not support more sophisticated dataflows such as row-stationary dataflow and quantization-aware lightweight processing elements which can enlarge the hardware accelerator design space and push the Pareto-frontier even further in terms of accuracy and hardware efficiency metrics such as energy-efficiency and performance per area. Moreover, this framework does not have a pre-characterized analytical hardware cost model. Therefore, it is not suitable for a rapid design space exploration of DNN accelerators and DNN accelerator and model co-exploration research. 
Therefore, there is a need for a framework that can assist system architects and machine learning researchers to quickly iterate over various hardware accelerator designs and DNN configurations while having an accurate hardware cost model and a flexible implementation to enable researchers to easily build their novel ideas on top of it without going through the tedious and laborious effort of RTL implementation from scratch. 

To this end, we propose \textit{QUIDAM}, a highly parameterized spatial-array based DNN accelerator framework that has implicit optimizations for CNNs as it is a domain-specific architecture but also has sufficient flexibility in terms of changing the microarchitectural features of the architecture, enriching the design space by providing lightweight processing element implementations, and crucial DNN model parameters such as bit precision to fully support DNN accelerator and model co-exploration.

As the systems community investigates novel hardware architectures and tools to enable deployment of efficient inference on edge, the ML community has focused on model compression techniques to achieve these objectives. Specifically, pruning~\cite{han2015deep_compression} and quantization~\cite{zhou2016dorefa} have received great interest in the ML community to design models for edge devices. In quantization, prior work focuses on improving fixed-point quantization~\cite{jacob2018quantization,esser2019learned}. Additionally, hardware-friendly quantization schemes have also been proposed~\cite{ruizhou2018lightnn,tambe2020algorithm,li2019additive}. In this work, we implement a specific quantization scheme chosen for its hardware-efficiency as it relies on reduced representations using a limited sum of power-of-two~\cite{ruizhou2018lightnn}.

\section{Methodology}\label{sec:quidam_methodology}

\begin{figure}[t]
  \centering
 \includegraphics[width=1\textwidth]{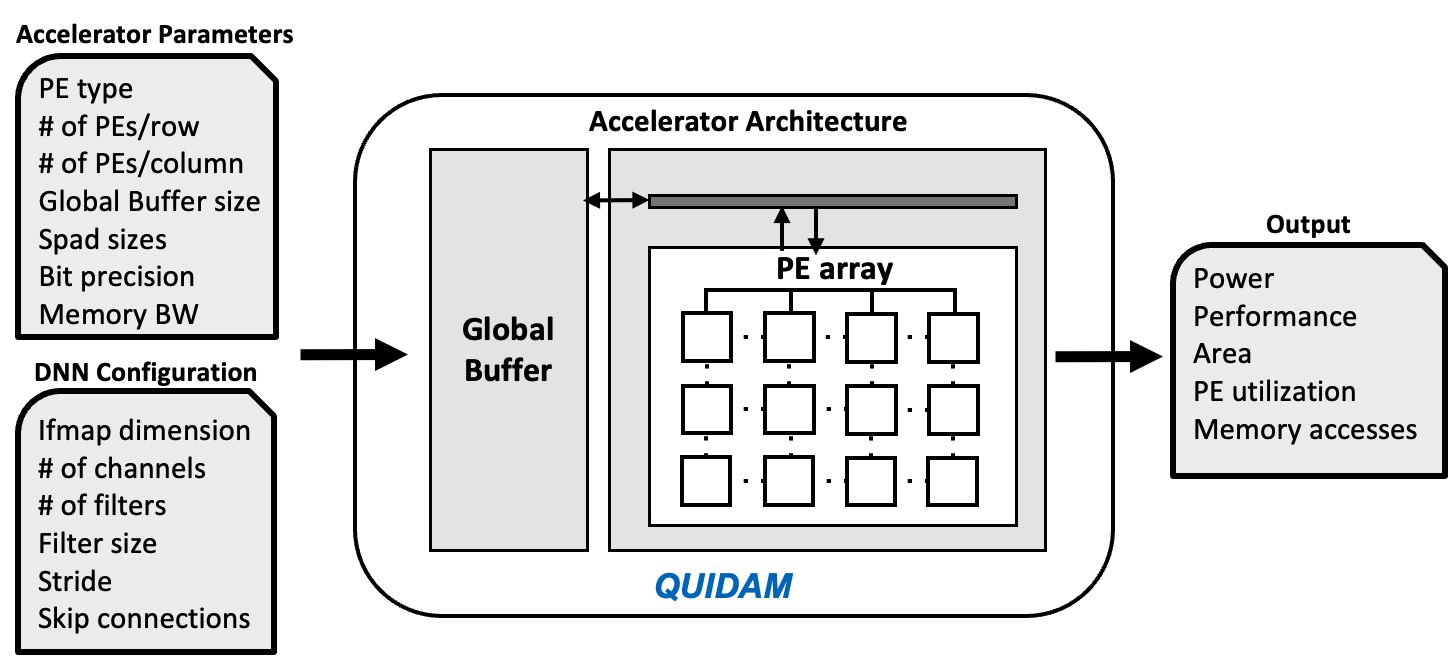}
 \caption{{Schematic depicting \textit{QUIDAM} framework, with accelerator parameters and DNN configuration as inputs. The framework takes in accelerator parameters and layer-wise DNN configurations and generates power, performance, area results, and statistics on hardware utilization and memory accesses.}}
 \label{fig:qappa}
\end{figure}

In this section, we present the proposed \textit{QUIDAM} framework, as shown in Figure~\ref{fig:flow}. First, the proposed \textit{QUIDAM} framework takes hardware accelerator parameters and deep neural network configurations as inputs to obtain power, performance, and area models for the next stages. Figure~\ref{fig:qappa} shows the available hardware accelerator and DNN configuration parameters that can be chosen using the \textit{QUIDAM} framework. We show the implementation details and architectural components of our \textit{QUIDAM} framework, as depicted in Figure~\ref{fig:qappa} (Section \ref{sec:quidam_framework}). Moreover, we also detail the lightweight processing elements (LightPE) that we implemented in our framework to provide a specialized processing element (PE) type for quantized DNN models (Section \ref{sec:quidam_lightpe}). After developing the power, performance, and area models for the \textit{QUIDAM} framework for a wide range of accelerator and DNN configurations (Section \ref{sec:quidam_modeling}), we perform a Pareto-optimality analysis for accuracy and crucial hardware efficiency metrics such as performance per area and energy (Section \ref{sec:quidam_pareto_perfarea}-\ref{sec:quidam_pareto_energy}). Finally, we carry out a DNN accelerator and model co-exploration analysis by generating candidate hardware and DNN configurations and demonstrate the generalizability of the proposed \textit{QUIDAM} framework by co-exploring the design space of both hardware and DNN configurations (Section \ref{sec:quidam_coexploration}). Therefore, the proposed framework provides power, performance, area, and RTL implementation using the developed power, performance, and area models and parameterized RTL implementation. Our proposed framework fosters future research on design space co-exploration of DNN accelerators and DNN configurations.

\subsection{QUIDAM Framework}\label{sec:quidam_framework}

To enable comprehensive design space exploration for DNN accelerators for on-device machine learning, we implemented \textit{QUIDAM}, a highly parameterized spatial-array based DNN accelerator framework in RTL. Our framework enables hardware designers and machine learning practitioners to rapidly iterate over various accelerator designs and DNN configurations and better understand trade-offs of different architectural components of the design for dizzying requirements of deploying machine learning models to edge devices. Moreover, hardware designers can also use the automatically generated RTL code to follow the design synthesis flow. 

As depicted in Figure~\ref{fig:qappa}, \textit{QUIDAM} framework is based on spatial-array based accelerators and utilizes row stationary dataflow which has been demonstrated to optimize the data movement in the storage hierarchy and improve the energy-efficiency of the system \cite{eyeriss}. \textit{QUIDAM} features a set of processing elements organized as a 2D array and a global buffer that stores input feature maps, filters, and activations. The number of PEs in each dimension can be tuned for different power, performance, and area requirements. Each PE includes an input feature map, a filter, partial sum scratchpads, and a multiply-accumulate (MAC) unit which can be implemented as a conventional MAC unit or a specialized shift-add unit based on the desired bit precision. Each of these architectural components can be tuned in a flexible and automated manner to perform a comprehensive design space co-exploration for on-device edge accelerators and DNN models.  
q

\subsection{Lightweight Processing Elements (LightPE)}\label{sec:quidam_lightpe}

To enrich the design space of hardware accelerators and achieve a better Pareto-frontier for performance per area and energy-efficiency, we include LightPE implementations in our framework. LightPEs utilize 8 bits for activations, as well as 4 bits and 8 bits for weights for LightPE-1 and LightPE-2 designs, respectively. As 4 bit and 8 bit quantization techniques for on-device machine learning have become prevalent in various computing platforms, we provide these specialized quantization-aware PE types in our \textit{QUIDAM} framework to help hardware designers to enrich their design space and find better Pareto-frontiers.

Specifically, LightPEs use a special power-of-two quantization scheme~\cite{ruizhou2018lightnn} that quantizes the weights of the neural network into a limited sum of powers-of-two. In this case, the multiplication between the activation and weight can then be replaced by shifts and adds. More generally speaking, a multiplication between an 8-bit activation $x$ and an 8-bit weight $w$ can be formulated as follows:
\begin{align}\label{eq:lightnn}
    \begin{split}
        y &= x\times w\\
        &= \sum_{i=0}^7 \mathds{1}(w_i) \times (x << i)\\
        &\text{where}~\mathds{1}(w_i) =
        \begin{cases}
              1 & \text{the $i^{th}$ bit of $w$ is 1}\\
              0 & \text{otherwise}
        \end{cases},
    \end{split}
\end{align}
where $<<$ denotes a left shift operator. Based on equation~\ref{eq:lightnn}, LightPE implementations approximate the sum with $k$ shifts and $k-1$ add operations, which was shown to be effective for significantly increasing energy-efficiency \cite{ruizhou2018lightnn,ruizhou2018journal}. We implement such an idea in LightPE-1 (for 1 shift) and LightPE-2 (for 2 shifts and one addition). In addition to LightPEs, we have also incorporated the design of a conventional 16 bit integer quantization (INT16) implementation and a conventional full-precision 32-bit floating point (FP32) implementation.

\begin{figure}
  \begin{subfigure}{0.49\textwidth}
    \includegraphics[width=\linewidth]{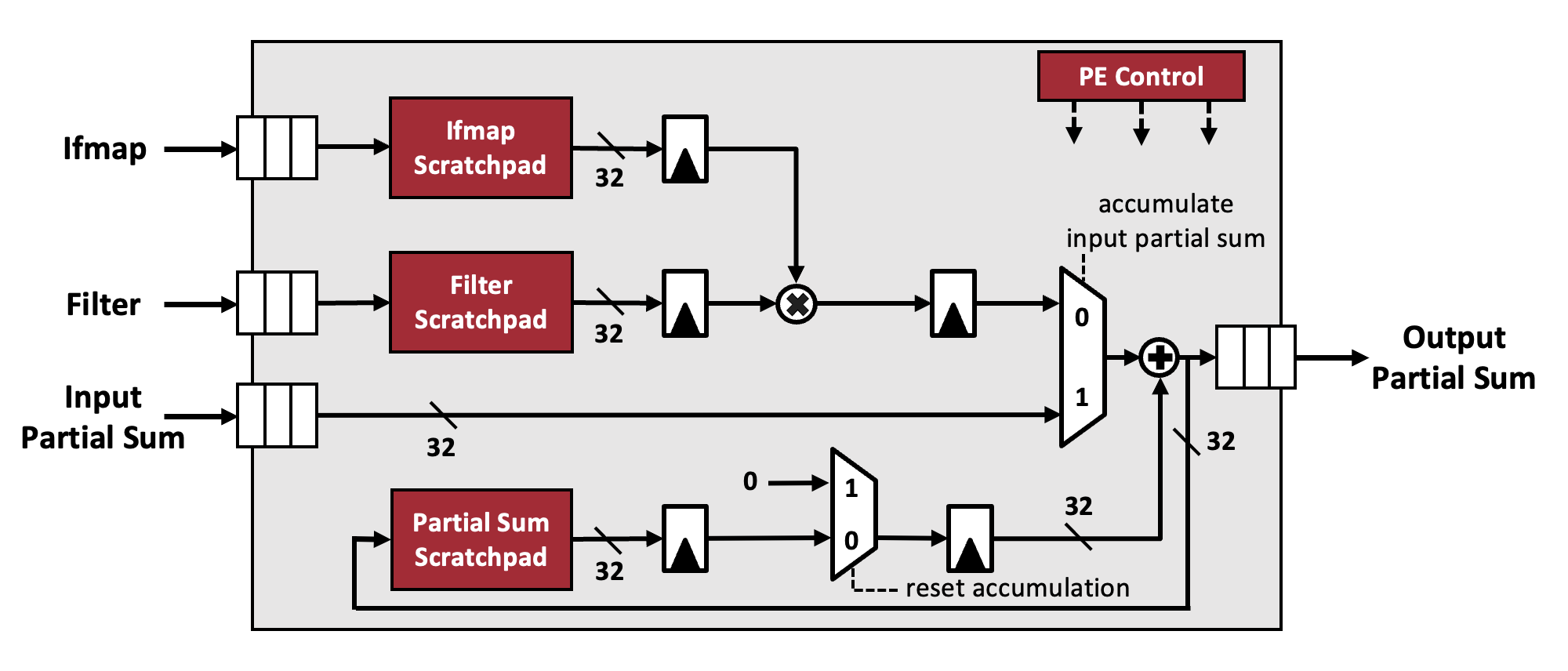}
    \caption{FP32 Processing Element} \label{fig:1a}
  \end{subfigure}%
  \hspace*{\fill}   % maximize separation between the subfigures
  \begin{subfigure}{0.49\textwidth}
    \includegraphics[width=\linewidth]{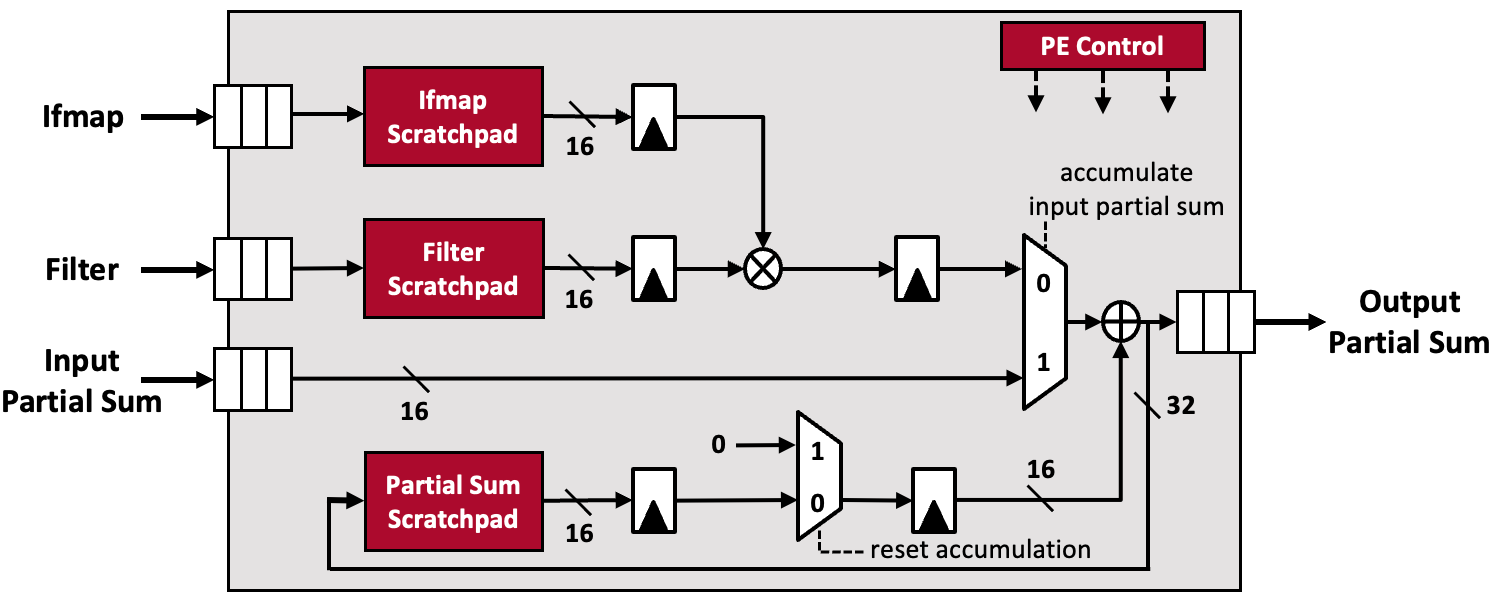}
    \caption{INT16 Processing Element} \label{fig:1b}
  \end{subfigure}%
  \hspace*{\fill}   % maximizeseparation between the subfigures
  \\
  \begin{subfigure}{0.49\textwidth}
    \includegraphics[width=\linewidth]{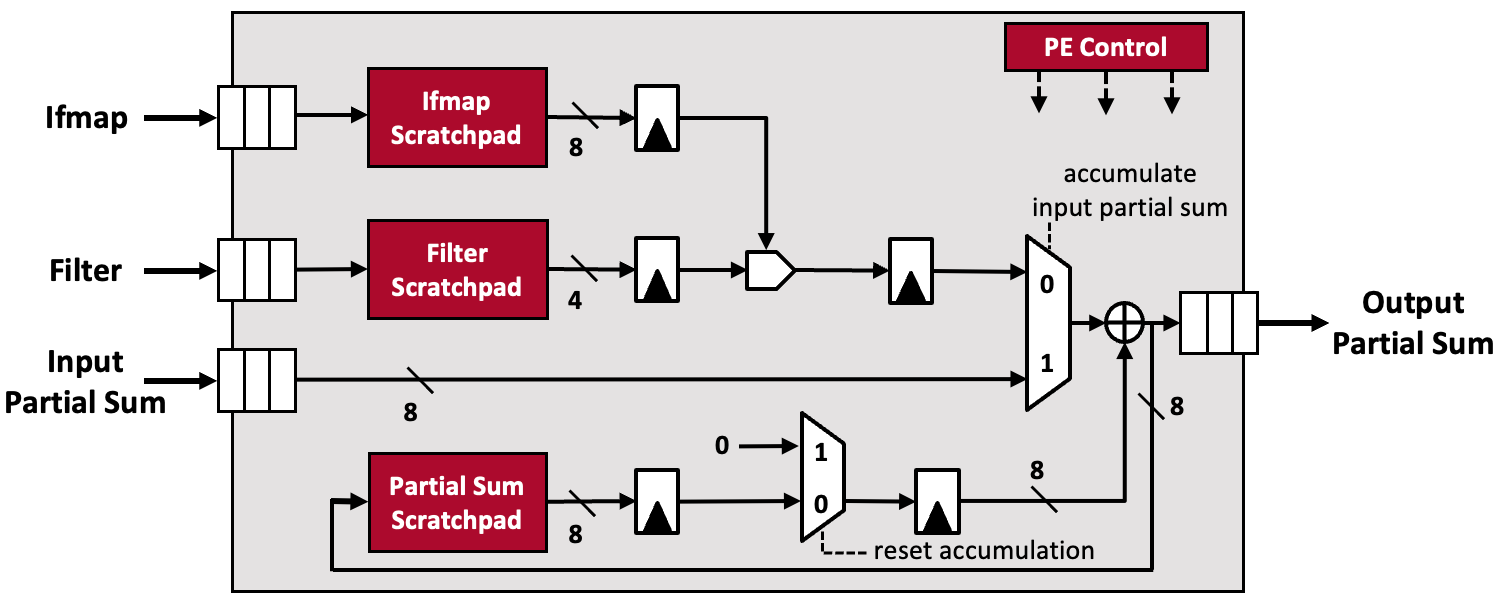}
    \caption{LightPE-1 Processing Element} \label{fig:1c}
  \end{subfigure}
    \hspace*{\fill}   % maximizeseparation between the subfigures
  \begin{subfigure}{0.49\textwidth}
    \includegraphics[width=\linewidth]{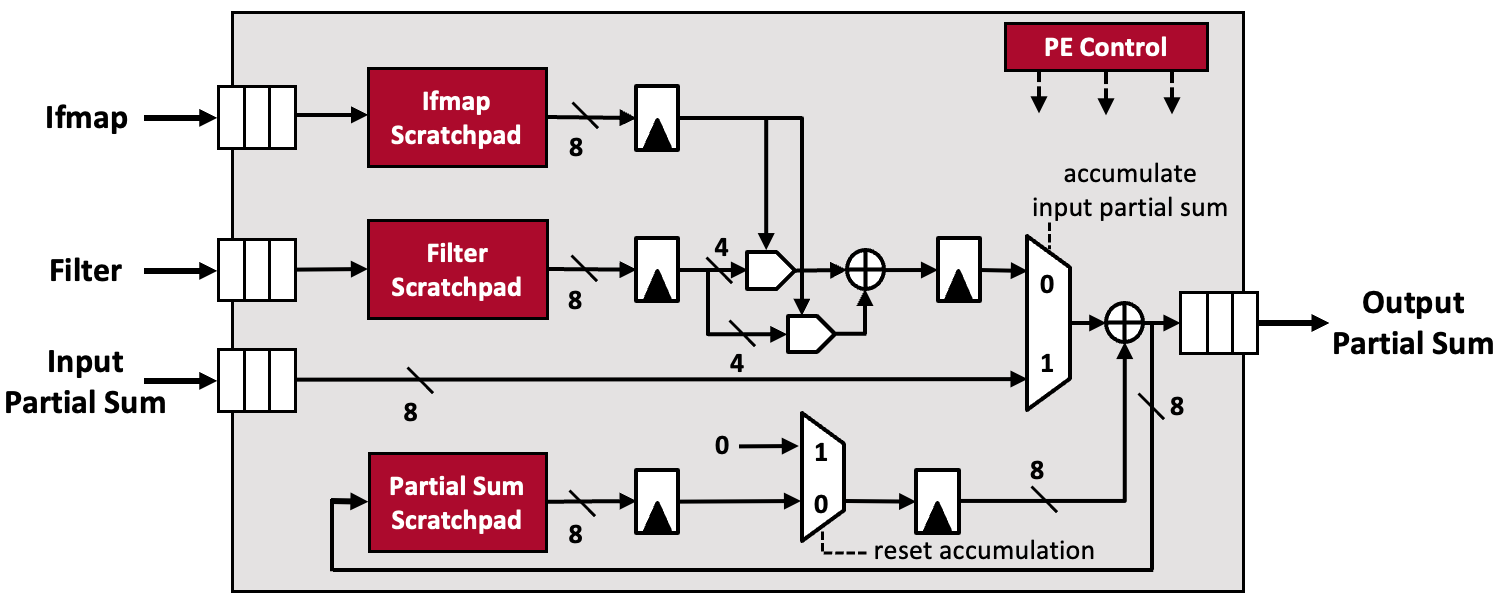}
    \caption{LightPE-2 Processing Element} \label{fig:1d}
  \end{subfigure}
  \\
    \begin{subfigure}{1\textwidth}
    \includegraphics[width=1\linewidth]{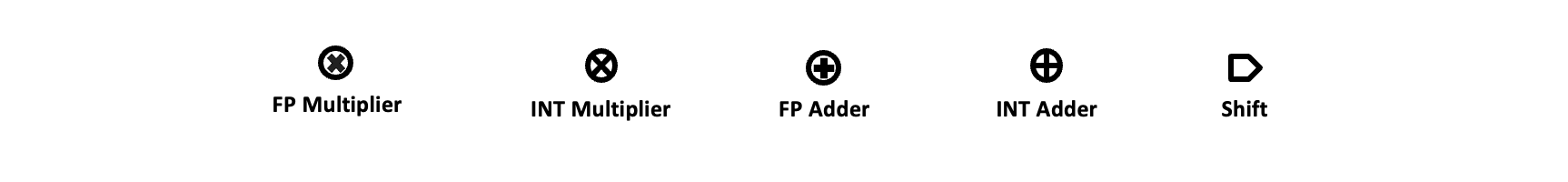}
  \end{subfigure}%
  \hspace*{\fill}    

\caption{Detailed architectures of FP32, INT16, LightPE-1, and LightPE-2 processing elements.}
\label{fig:pe_element}
\end{figure}

Figure \ref{fig:pe_element} shows the detailed architectures of FP32, INT16, LightPE-1, and LightPE-2 processing elements. Each processing element has four FIFOs for input feature map (ifmap), filter, input partial sum, and output partial sum. Moreover, there are three scratchpad memories implemented in each processing element such as input feature map scratchpad, filter scratchpad, and partial sum scratchpad which get the data from aforementioned FIFOs. After getting the data from scratchpads, there are different versions of multiplication implementations between weights and activations for different processing elements. More specifically, Figure \ref{fig:1a} shows the FP32 processing element that is used in \textit{QUIDAM} framework. The FP32 implementation uses a 32-bit floating point multiplier and adder as it is commonly implemented in conventional systems. Figure \ref{fig:1b} shows the INT16 processing element that has an 16-bit integer multiplier and adder. Figure \ref{fig:1c} and Figure \ref{fig:1d} show LightPE-1 and LightPE-2 implementations, respectively. The LightPE-1 implementation utilizes a shift instead of a multiplier and it uses 8 bits for activations and 4 bits for weights. To store a weight $w$ = $\pm 2^{-m}$, where $m$ = $0,1, ... , 7$, LightPE-1 needs four bits: one bit for the \textit{sign(w)} and three bits for $|m|$. On the other hand, the LightPE-2 implementation utilizes two shifts and an addition as shown in Figure \ref{fig:1d}. LightPE-2 utilizes 8 bits for activations and 8 bits for weights. To store a weight $w$ = $\pm (2^{-m_{1}} + 2^{-m_{2}})$, where $m_{1},m_{2}$ = $0, 1, ... , 7$, LightPE-2 requires seven bits: one bit for the \textit{sign(w)}, three bits for $|m_{1}|$, and three bits for $|m_{2}|$. For easier hardware implementation, 8 bits are used. After completing the multiplication and shift operations, all processing element implementations rely on a multiplexer for accumulating input partial sum data. Similarly, a second multiplexer is used for partial sum scratchpad to reset the accumulation. Finally, after the final addition operation, the data is sent to the output partial sum FIFO and the result is available.

\begin{figure*}[t]
  \centering
 \includegraphics[width=0.8\textwidth]{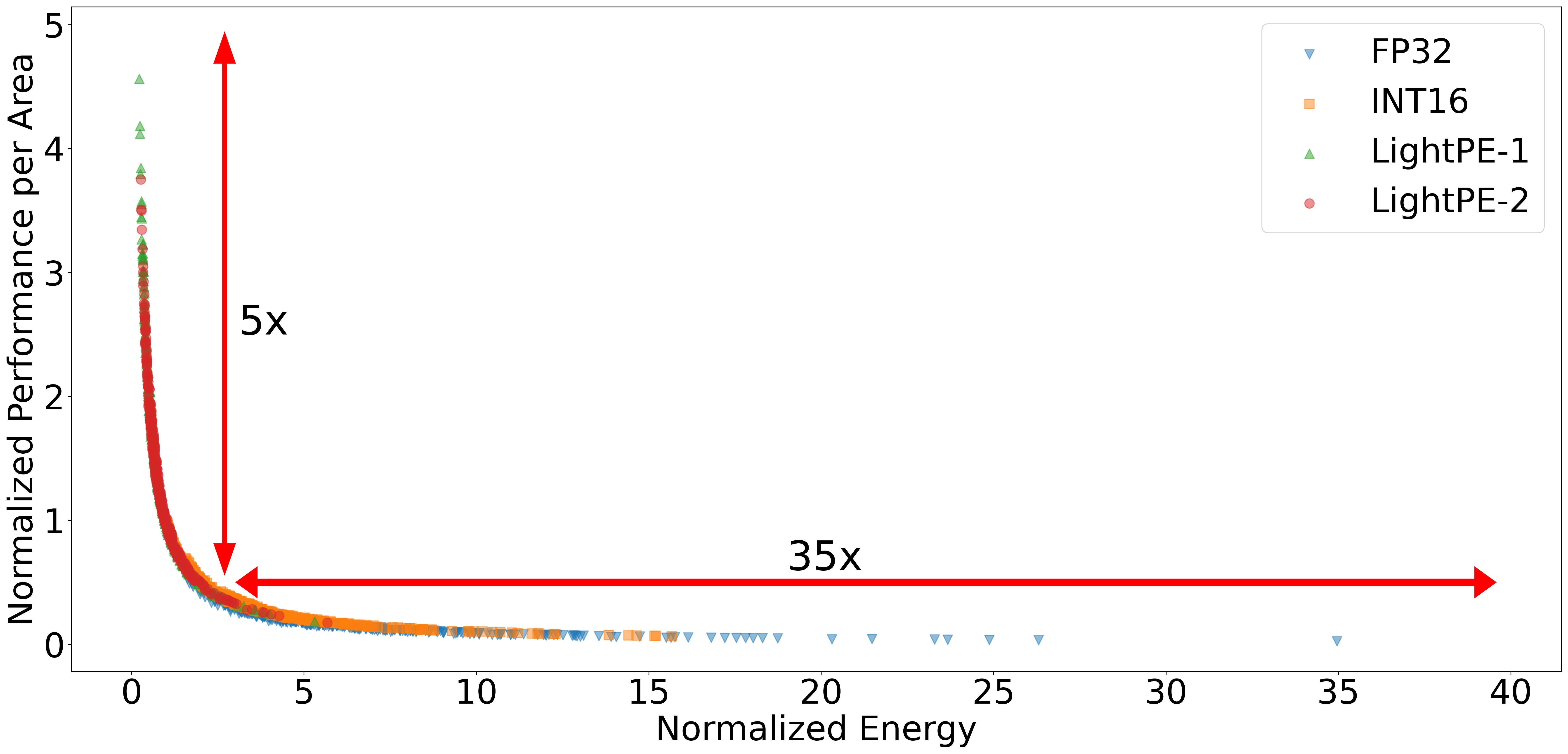} 
 \caption{{Different PE types and bit precision lead to significant differences in performance per area and energy. Therefore, there is a need for a design space exploration framework that incorporates quantization-aware processing elements and rapidly iterate over various designs.}}
\label{fig:divergence}
\end{figure*}

Besides their low-precision benefits such as reducing the storage requirements, LightPEs also replace the multiplications with a more energy and area-efficient shift operation or a limited number of shifts and add operations \cite{ruizhou2018lightnn,ruizhou2018journal}. Therefore, LightPEs also achieve significant energy and area gains when compared to full-precision 32-bit floating point and 16-bit integer based designs. As a result, LightPEs provide an enriched design space for hardware designers and machine learning practitioners to analyze various trade-offs between accuracy and performance per area and energy. 

To this end, we perform a design space exploration analysis with four different processing element types such as FP32, INT16, LightPE-1, and LightPE-2 and compare them in terms of normalized performance per area and normalized energy with respect to the INT16 design point with the highest performance per area. As seen in Figure \ref{fig:divergence}, normalized energy varies $35 \times$ for almost the same performance per area region and normalized performance per area varies $5 \times$ for almost the same energy region. In addition, while most of the configurations are clustered around the knee of the scatter plot regardless of the quantization level, we note that FP32 configurations dominate the highest energy ones, while LightPE-1 configurations are the ones that push performance per area to highest values, orders of magnitude larger than the INT16 case. Therefore, different PE types and precision levels lead to significant differences in terms of performance per area and energy. These results also reinforce the need for a design space exploration framework that incorporates quantization-aware hardware and rapidly iterates over various designs.

\subsection{Power, Performance, and Area Modeling}\label{sec:quidam_modeling}

To build our quantization-aware power, performance, and area models, we use various hardware and DNN configurations. Specifically, to cover this comprehensive design space of hardware accelerators, we generate a variety of possible designs by varying global buffer size, number of PEs per row and column in the 2D PE array, bit precision, and PE type (FP32, INT16, LightPE-1, and LightPE-2). Within each PE, we also vary individual scratchpad sizes for input feature map, filter, and partial sum scratchpads. 

We use Synopsys Design Compiler and the open-source FreePDK45 which is a commonly used process design kit \cite{freepdk} to synthesize our designs to obtain power, area, and initial timing results. We use Synopsys VCS RTL simulator to perform functional verification and collect timing information for various DNN configurations such as VGG-16 \cite{vgg16}, ResNet-20, ResNet-34, ResNet-50, and ResNet-56 \cite{resnet50} that are implemented in our testbenches. 
After collecting power, area, and timing results from these tools, we use polynomial regression models and model selection techniques based on \textit{k}-fold cross validation \cite{kfold} to tune the degree of the polynomial. 

More concretely, a $K$-degree polynomial regression model is defined as:

\begin{align}
    \begin{split}
        F(\bm{x}) = \sum_j c_j \prod x_i^{q_{ij}}\\
        \text{where}~\bm{x}\in \mathbb{R}^{d}; q_{ij}\in \mathbb{N}; \forall j, \sum_{i} q_{ij} \leq K,
    \end{split}
\end{align}
where $\bm{x}$ is a $d$-dimensional input feature vector with $x_i$ the $i^{\text{th}}$ feature, $j$ index denotes the $j^{\text{th}}$-term of the polynomial, and $\forall j,~c_j$ are learnable coefficients. We note that using polynomial regression in modeling the power and performance characterizing deep neural networks when running on a fixed hardware (\textit{i.e.}, desktop GPUs) has been studied before~\cite{cai2017neuralpower,qi17paleo}. To this end, for the novel design space where both the hardware and network configurations can vary, we explain in the sequel how features should be chosen for modeling. We detail the feature space for our proposed power, area, and latency models in the following.

\textbf{Power}\space\space\space We use Synopsys Design Compiler with inherently assumed switching activity when generating the power consumption. As a result, we choose $\bm{x}$ to be a four dimensional vector that includes: the scratchpad size for input feature map ($\mathsf{SP_{\mathsf{if}}}$), the scratchpad size for partial sum ($\mathsf{SP_{\mathsf{ps}}}$), the scratchpad size for filter weights ($\mathsf{SP_{\mathsf{fw}}}$), and the number of PEs ($\mathsf{\#PE}$). Finally, we develop individual models for each processing element type to improve the performance of our models as power depends on the PE type.

\textbf{Area}\space\space\space For area modeling, we use the same features as in power modeling because the features that affect power and area come from the same source. The area model only depends on the hardware configuration in contrast to the latency model which depends on both the hardware and the deep neural network configuration. More specifically, we choose $\bm{x}$ to be a four dimensional vector that includes: the scratchpad size for input feature map ($\mathsf{SP_{\mathsf{if}}}$), the scratchpad size for partial sum ($\mathsf{SP_{\mathsf{ps}}}$), the scratchpad size for filter weights ($\mathsf{SP_{\mathsf{fw}}}$), and the number of PEs ($\mathsf{\#PE}$). Similar to power modeling, we build individual models for each processing element type as the arithmetic units differ between PE types.

\begin{figure}
  \begin{subfigure}{0.49\textwidth}
    \includegraphics[width=\linewidth]{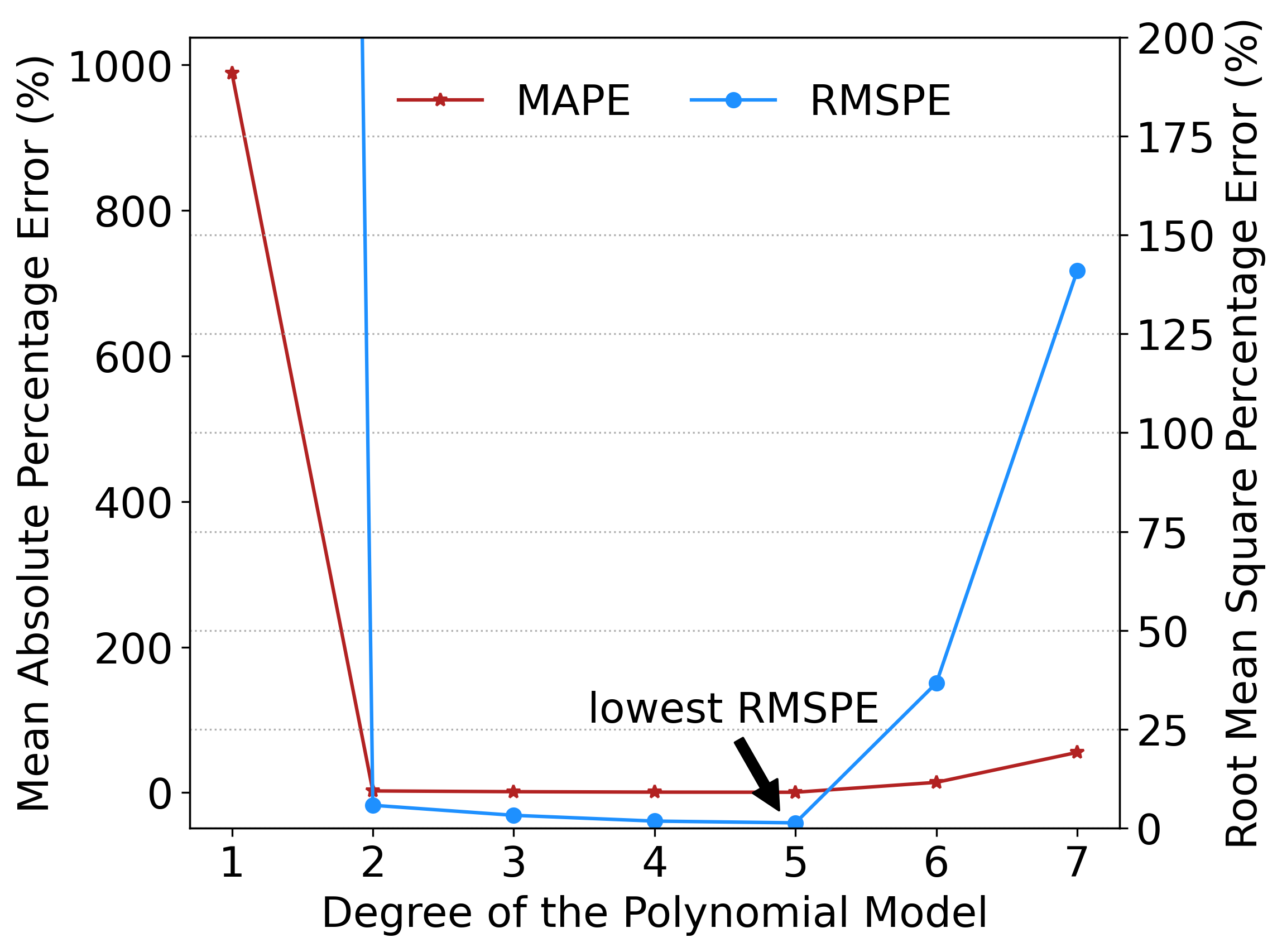}
    \caption{Power Modeling} \label{fig:power_model_selection}
  \end{subfigure}%
  \hspace*{\fill}   % maximize separation between the subfigures
  \begin{subfigure}{0.49\textwidth}
    \includegraphics[width=\linewidth]{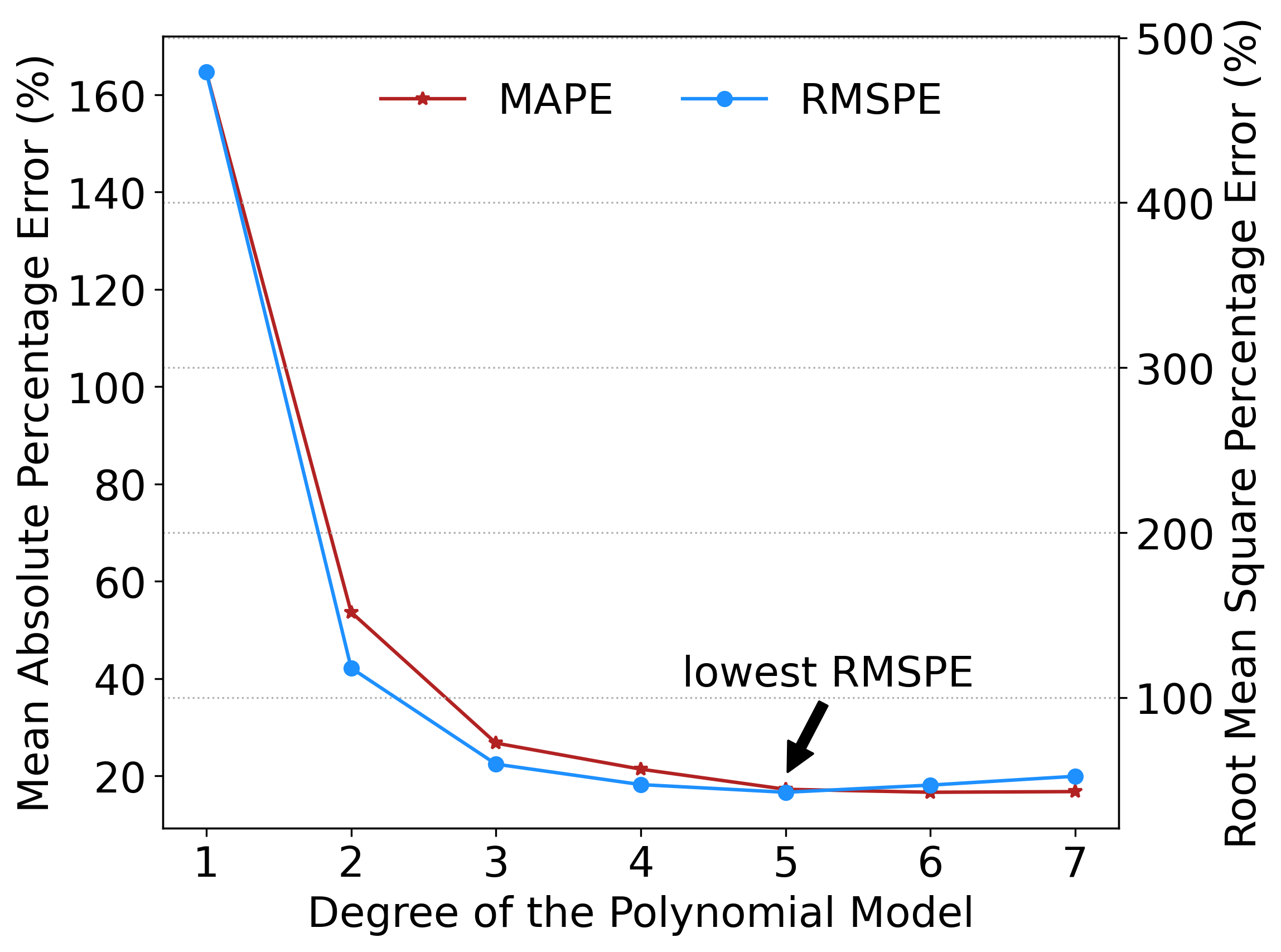}
    \caption{Performance Modeling} \label{fig:performance_model_selection}
  \end{subfigure}%
  \hspace*{\fill}   % maximizeseparation between the subfigures
  \\
  \hspace*{\fill}
  \begin{subfigure}{0.49\textwidth}
    \includegraphics[width=\linewidth]{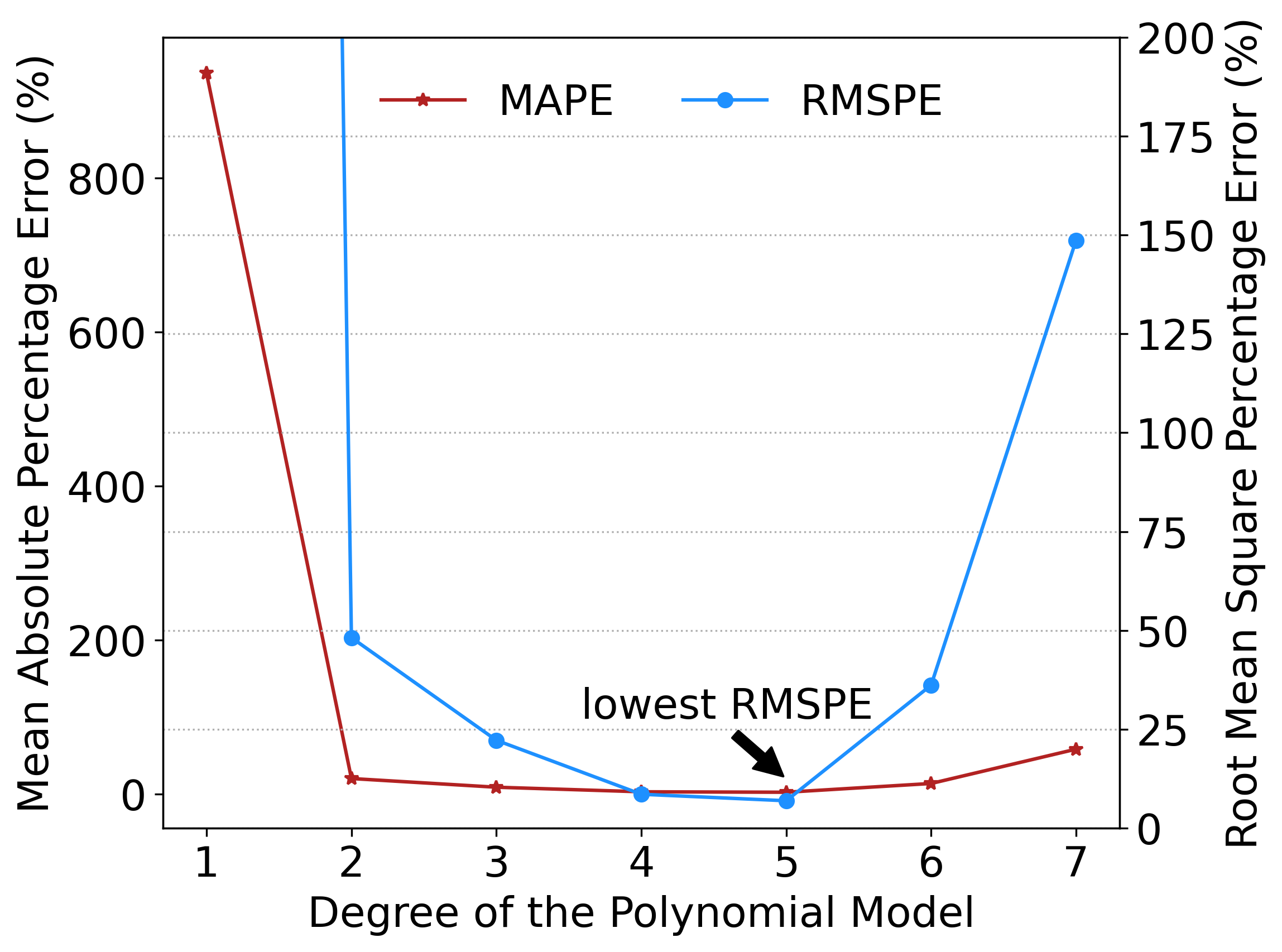}
    \caption{Area Modeling} \label{fig:area_model_selection}
  \end{subfigure}
    \hspace*{\fill}   % maximizeseparation between the subfigures

\caption{Comparison of the performance results of the model with respect to the degree of the polynomial model. A polynomial order of five is chosen for the power, performance, and area modeling since it achieves the lowest Root Mean Square Percentage Error (RMSPE) and Mean Absolute Percentage Error (MAPE) at the same time.}
\label{fig:model_selection}
\end{figure}

\begin{figure*}[t]
  \centering
 \includegraphics[width=0.49\textwidth]{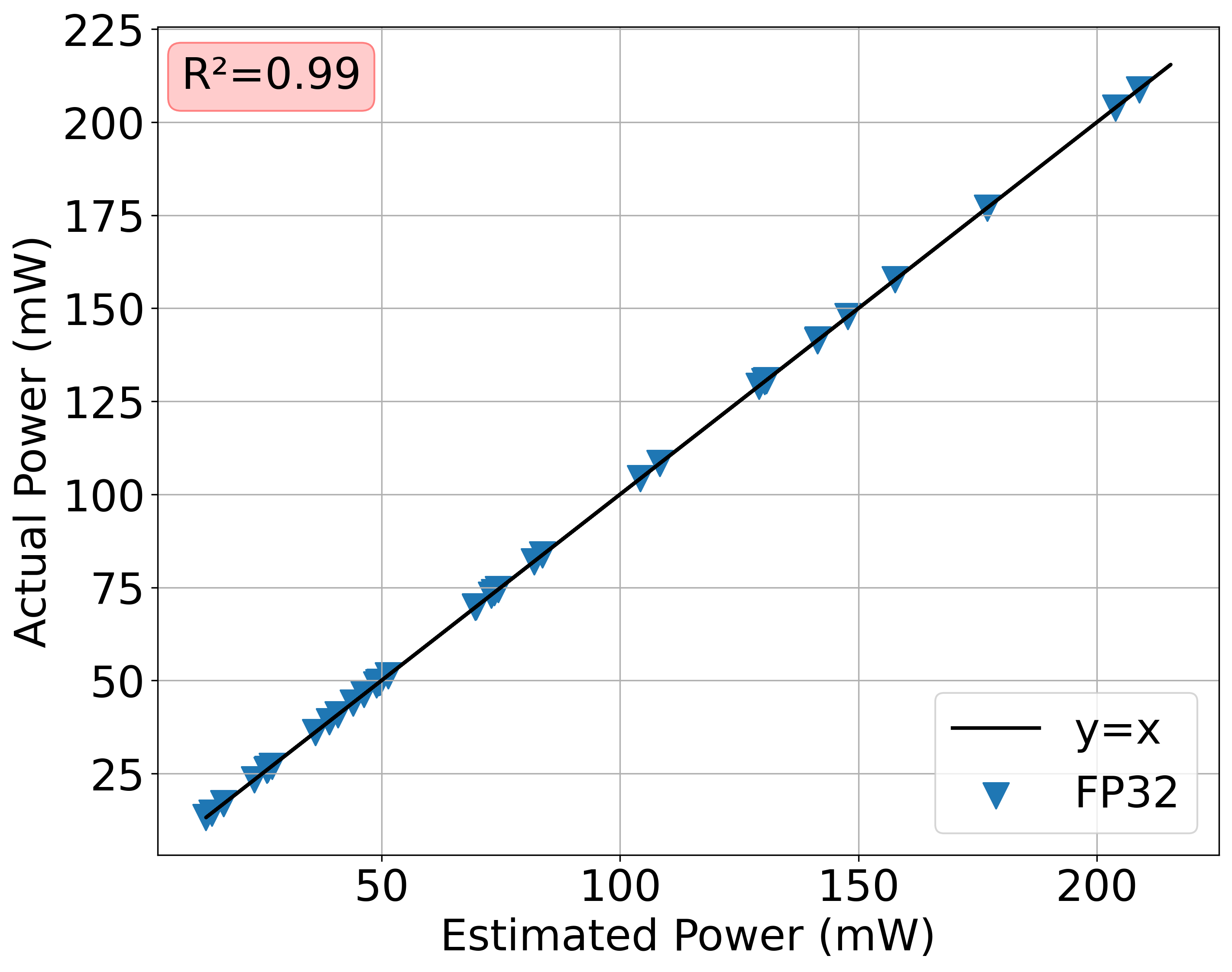}
  \includegraphics[width=0.49\textwidth]{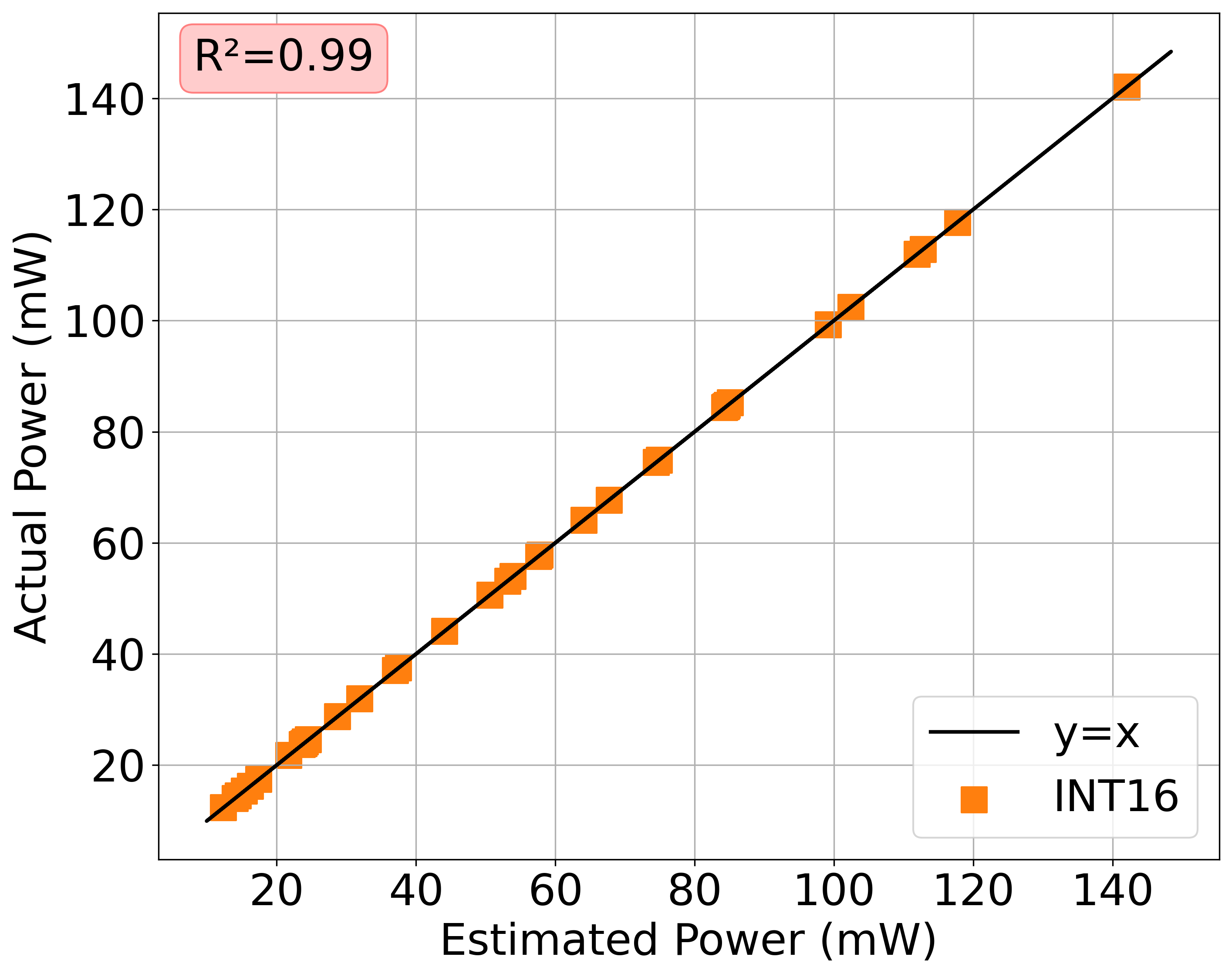}
  \\
  \includegraphics[width=0.49\textwidth]{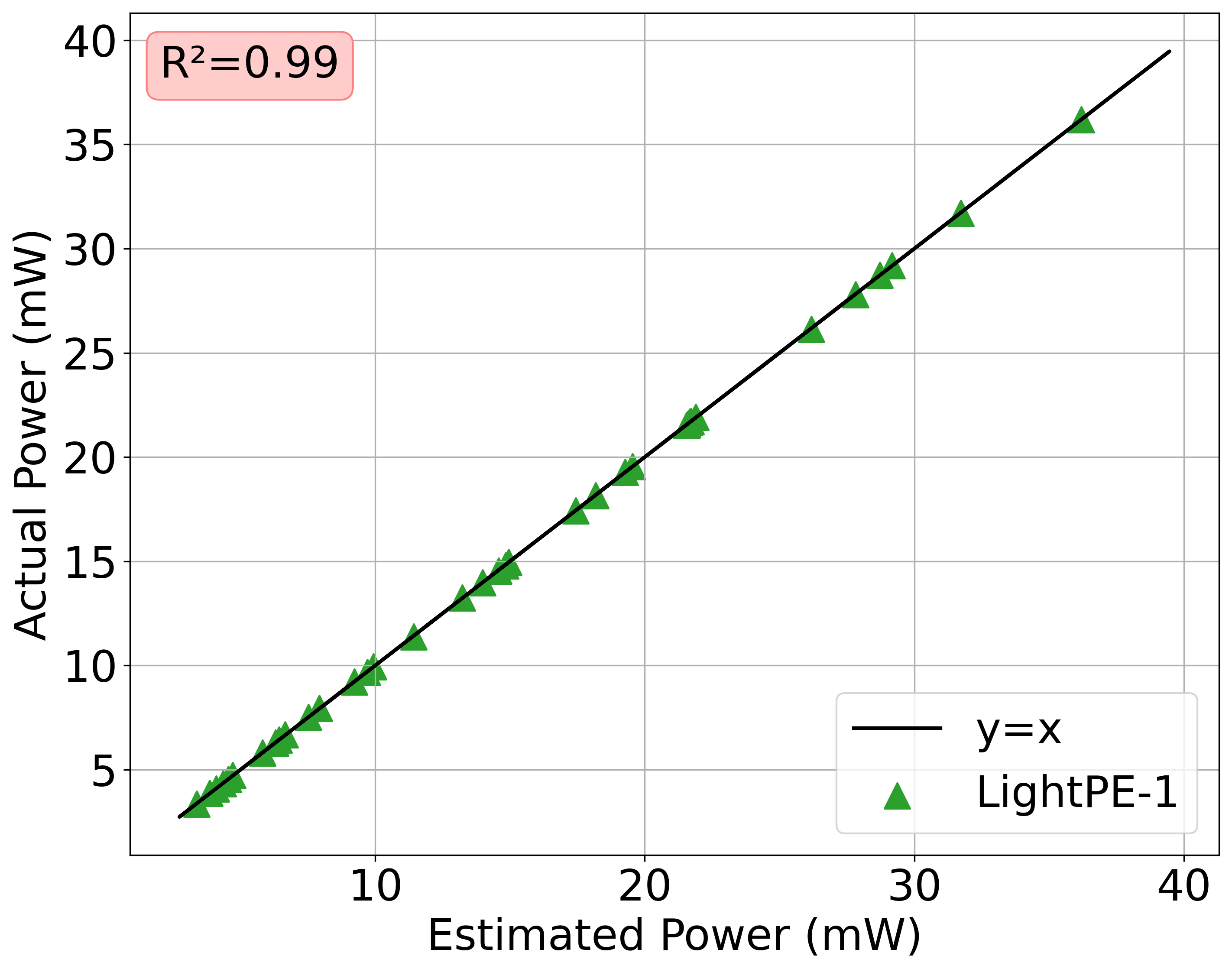}
    \includegraphics[width=0.49\textwidth]{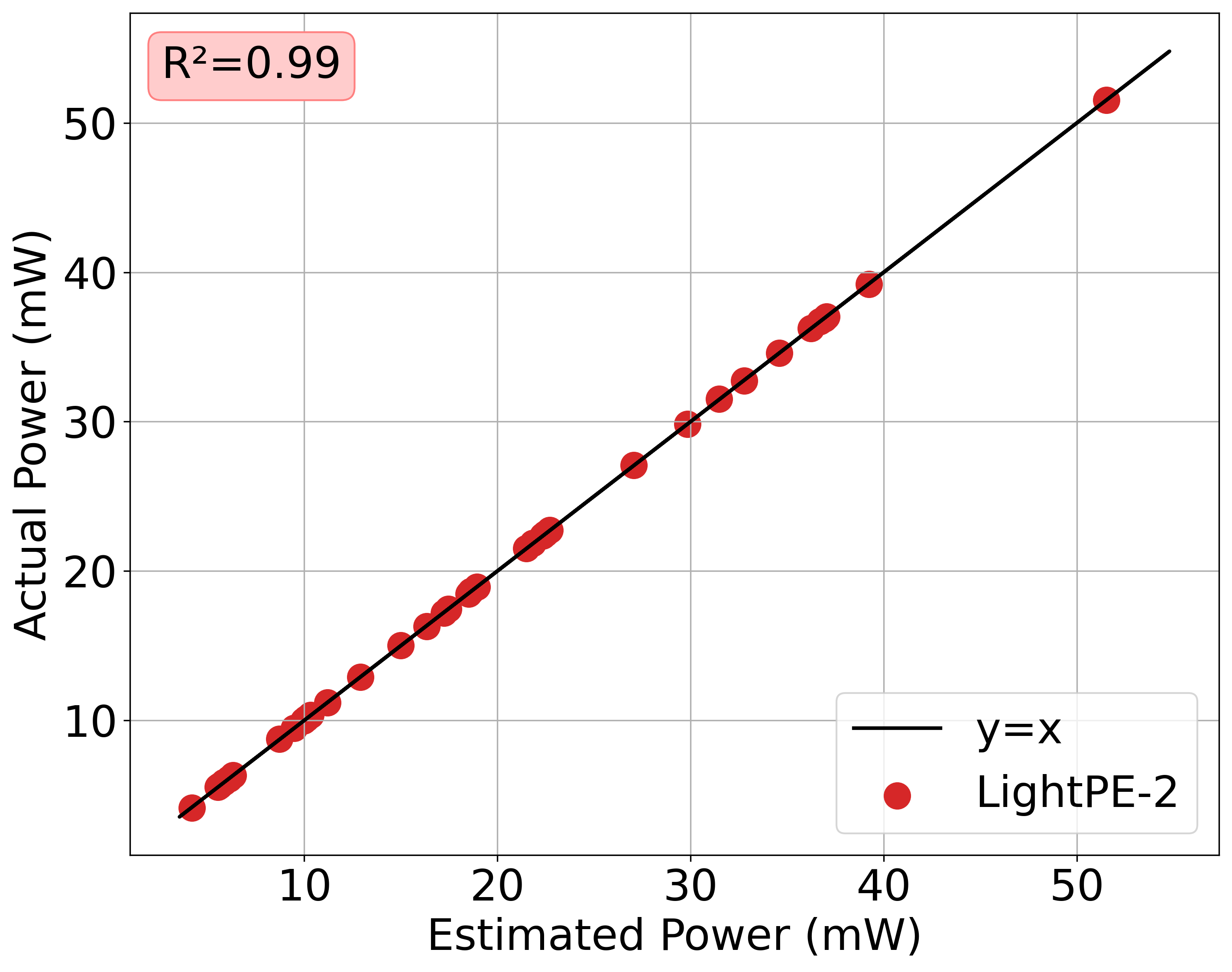}
    \caption{{Power estimation results for various processing element types such as FP32, INT16, LightPE-1, and LightPE-2. Each data point corresponds to a different hardware configuration that can be achieved by using the corresponding processing element type. As it can be seen, the proposed polynomial model agrees closely with the actual values extracted from the synthesis tools.}}
 \label{fig:power_model}
\end{figure*}

\begin{figure*}[t]
  \centering
       \includegraphics[width=0.49\textwidth]{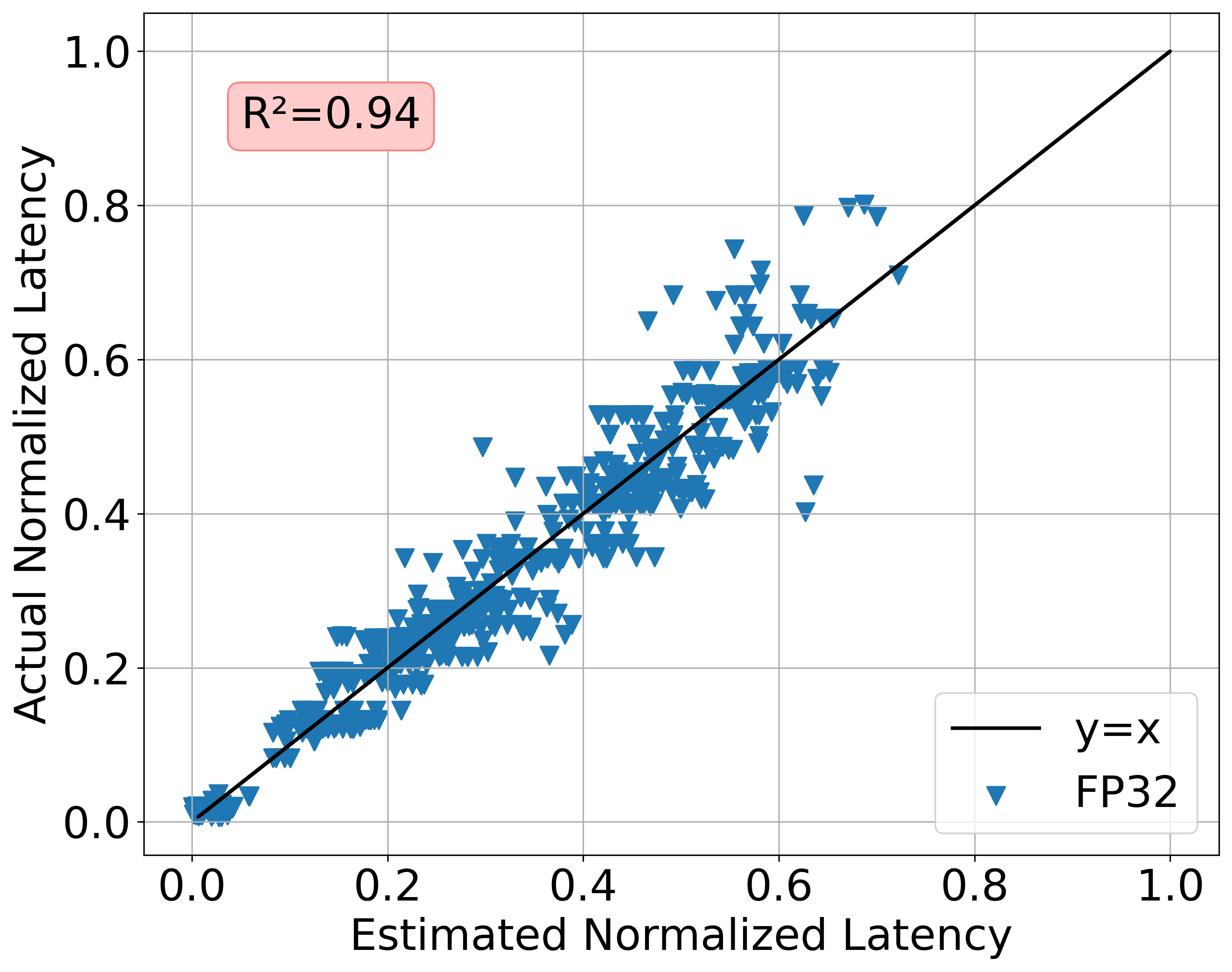} %fp32vgg
  \includegraphics[width=0.49\textwidth]{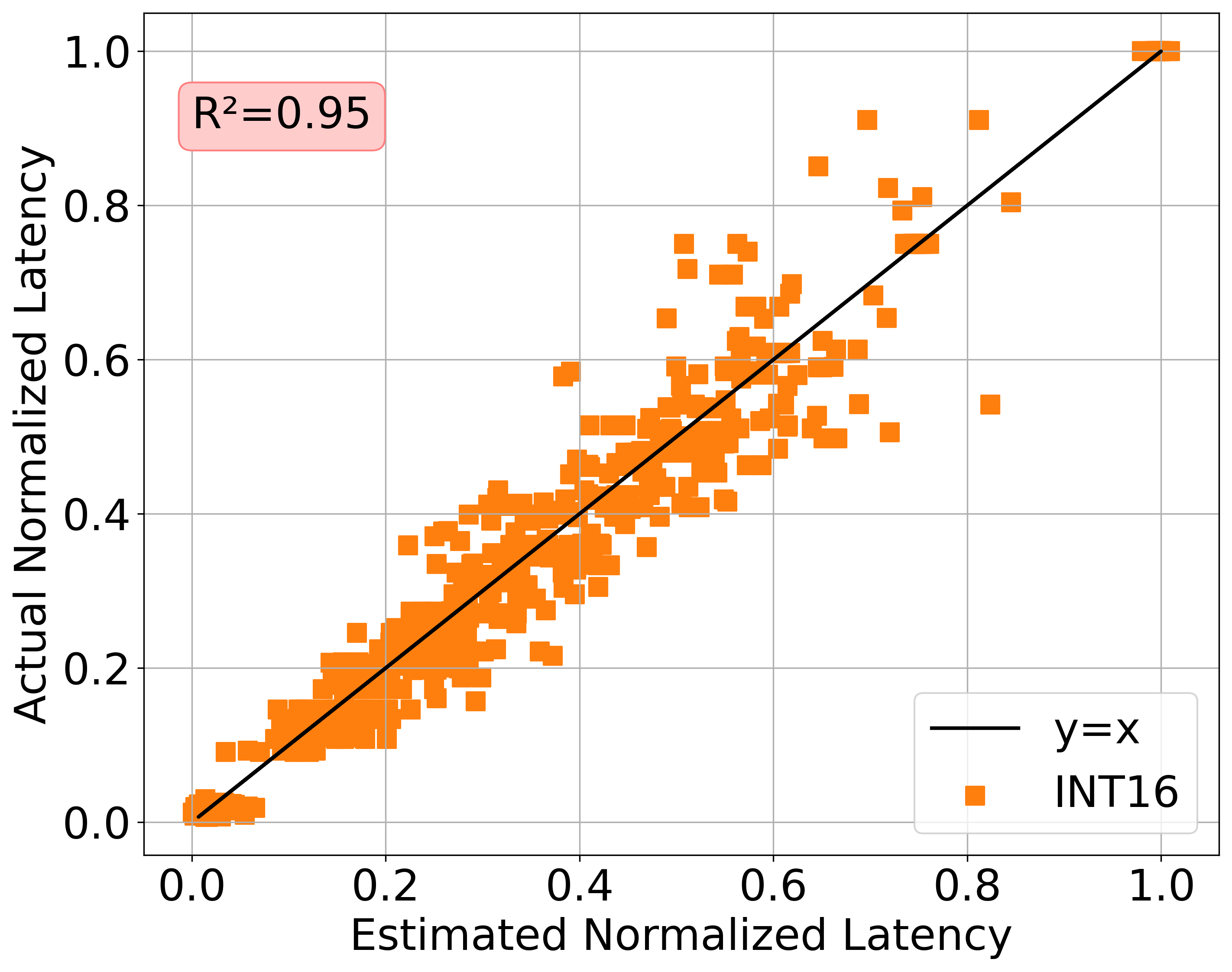}
  \\
  \includegraphics[width=0.49\textwidth]{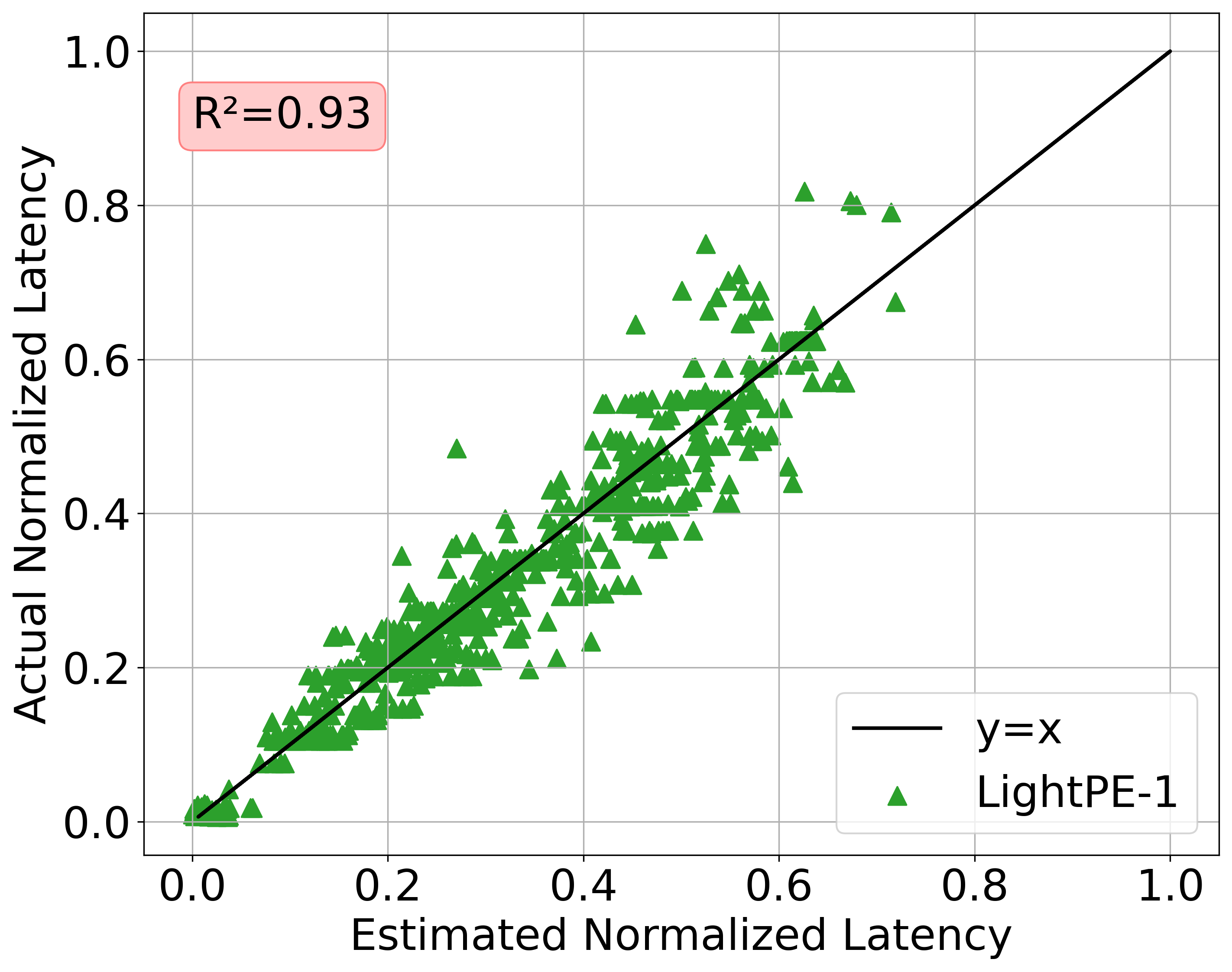}
    \includegraphics[width=0.49\textwidth]{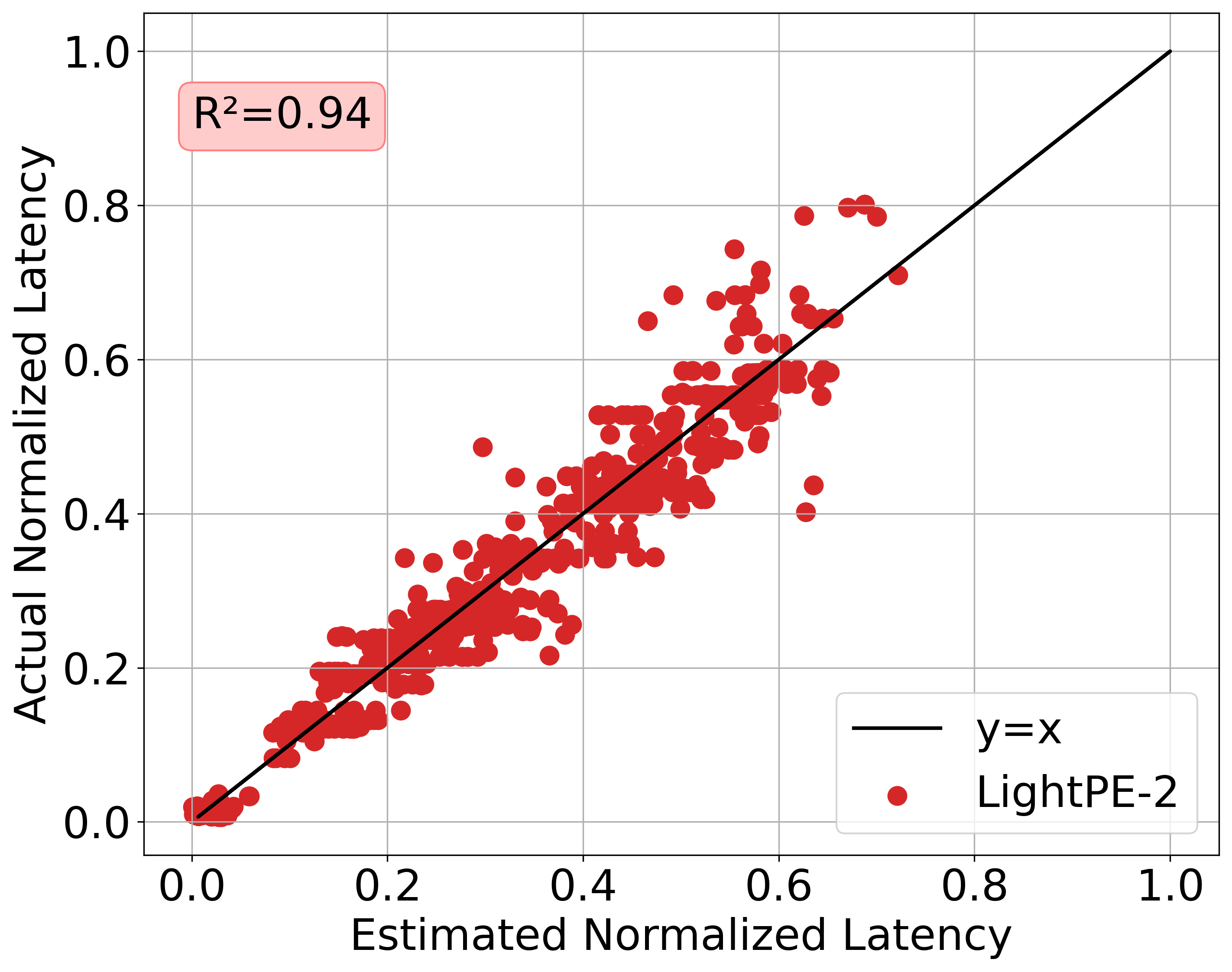}
    \caption{{Performance estimation results for various processing element types such as FP32, INT16, LightPE-1, and LightPE-2. Each data point corresponds to a different hardware configuration that can be achieved by using the corresponding processing element type. As it can be seen, the proposed polynomial model agrees closely with the actual values extracted from the synthesis tools.}}
%   \vspace{-2mm}
 \label{fig:performance_model}
\end{figure*}

\begin{figure*}[t]
  \centering
        \includegraphics[width=0.49\textwidth]{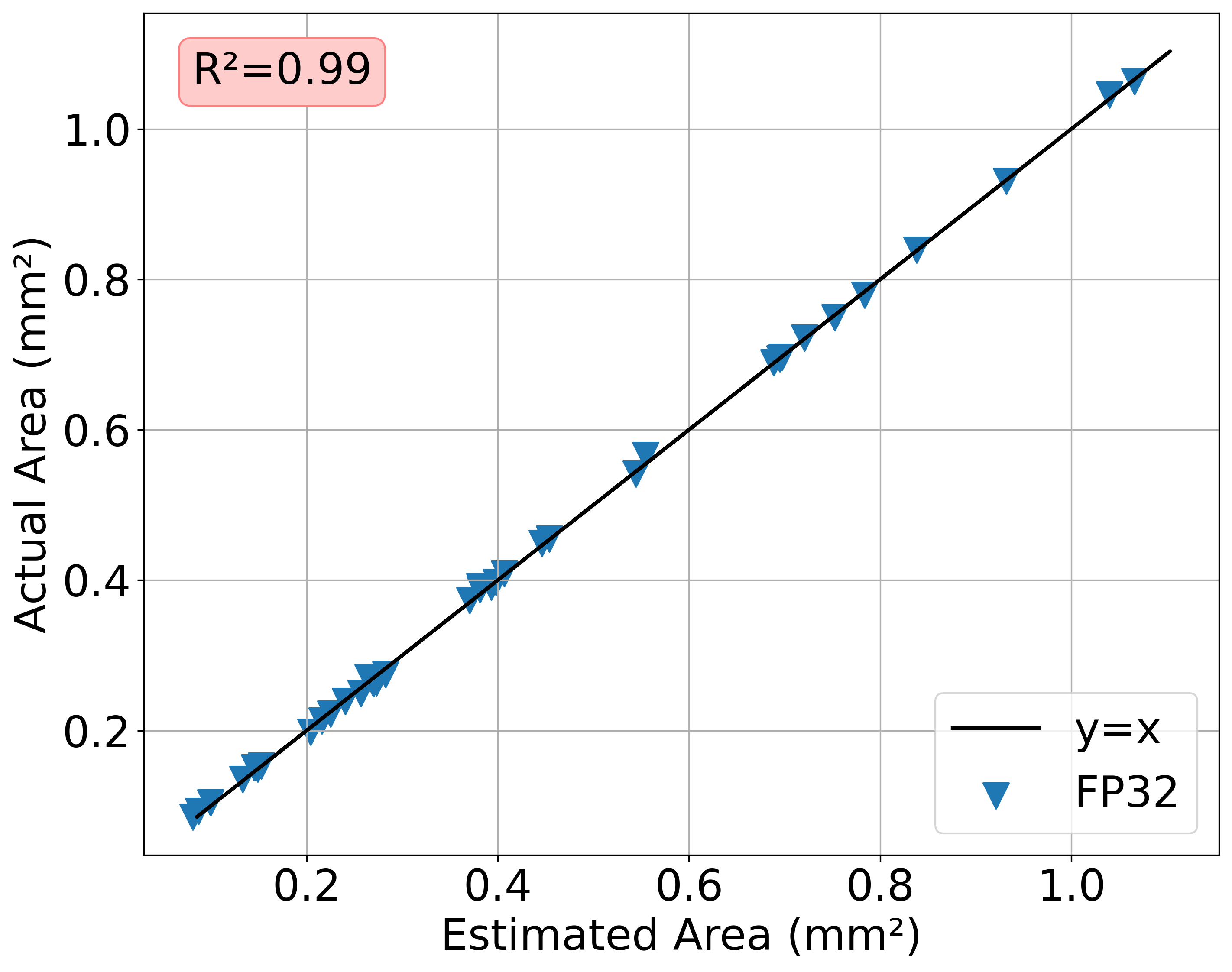}
  \includegraphics[width=0.49\textwidth]{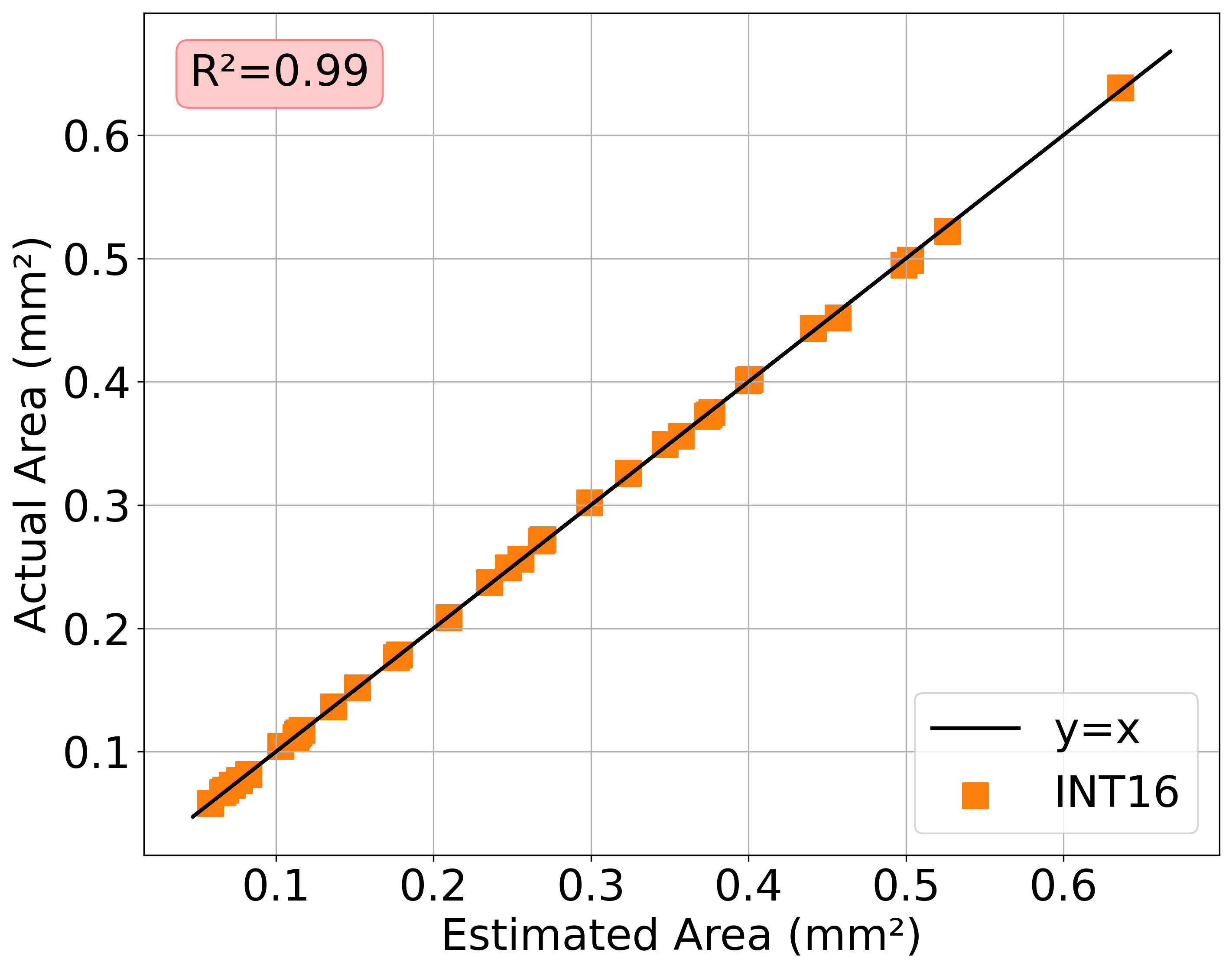}
  \\
  \includegraphics[width=0.49\textwidth]{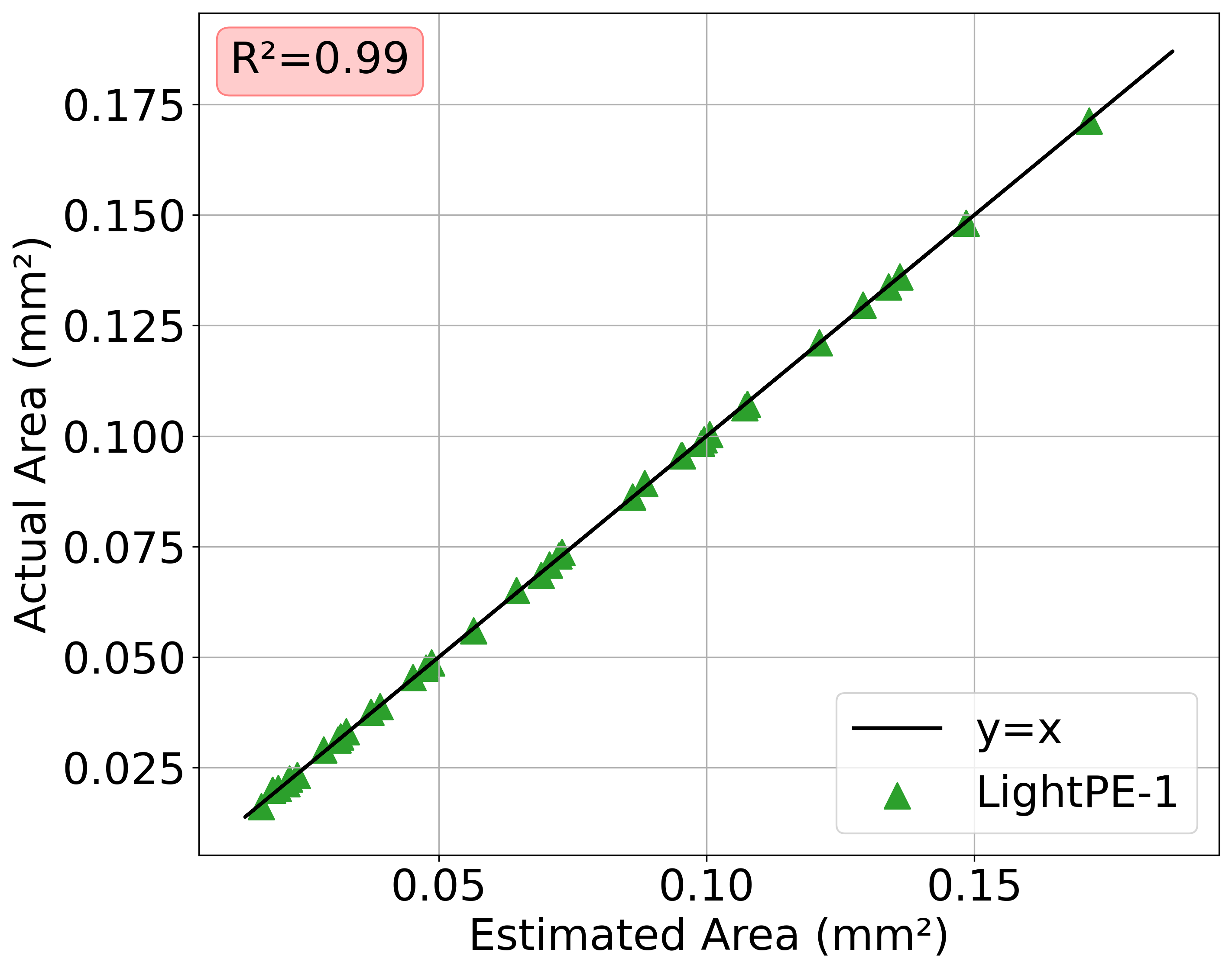}
    \includegraphics[width=0.49\textwidth]{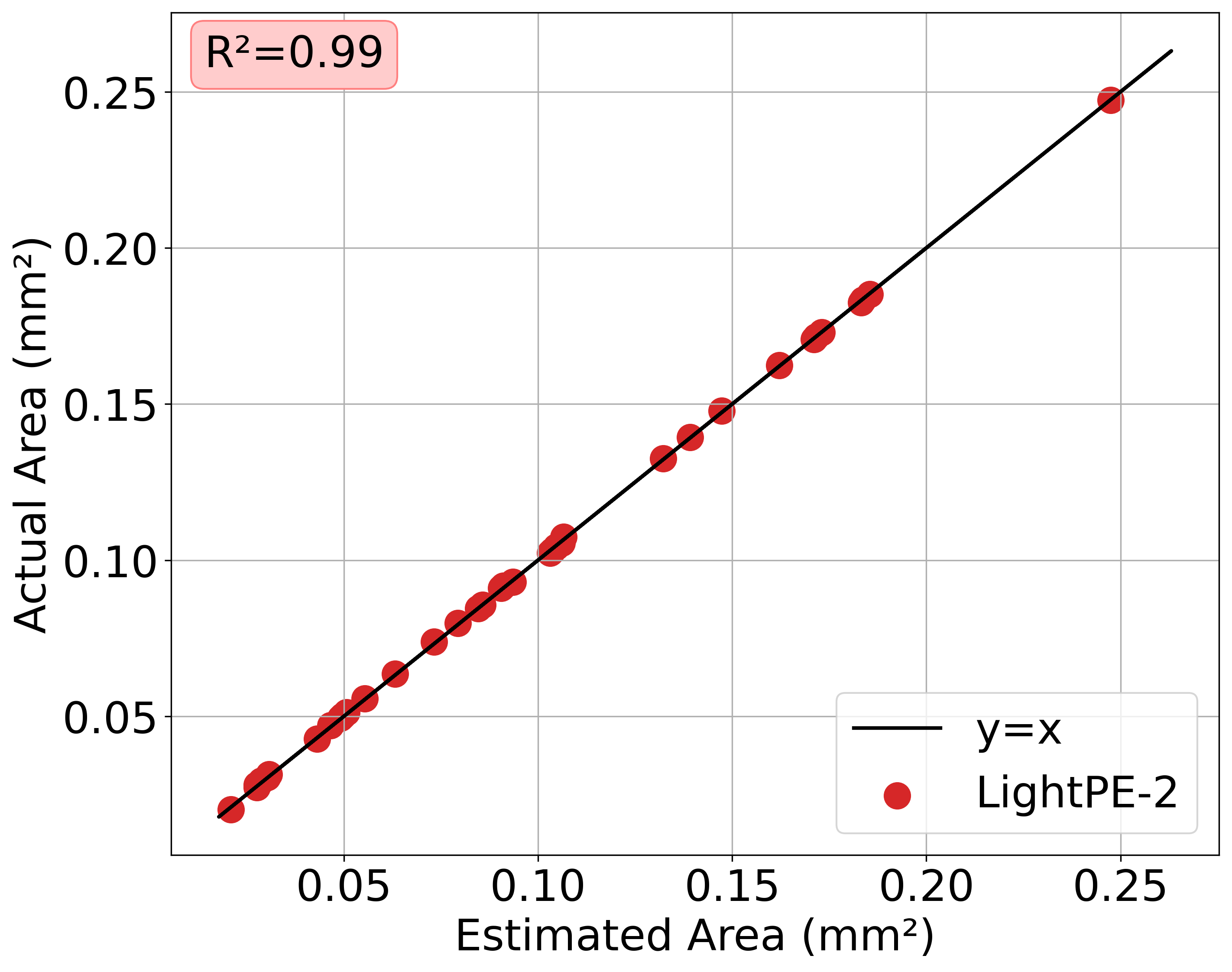}
    \caption{{Area estimation results for various processing element types such as FP32, INT16, LightPE-1, and LightPE-2. Each data point corresponds to a different hardware configuration that can be achieved by using the corresponding processing element type. As it can be seen, the proposed polynomial model agrees closely with the actual values extracted from the synthesis tools.}}
%   \vspace{-2mm}
 \label{fig:area_model}
\end{figure*}

\textbf{Latency}\space\space\space Latency depends on both the hardware and the neural network configurations. To cope with diverse network configurations, we adopt a layer-level latency modeling strategy. Specifically, we use the polynomial model to infer the per-layer latencies and sum them to obtain a network-level value. As a result, our training data for the polynomial model is at a layer-level granularity. We use a 12-dimensional feature vector for the latency model, which includes $\mathsf{SP_{\mathsf{if}}}$, $\mathsf{SP_{\mathsf{ps}}}$, $\mathsf{SP_{\mathsf{fw}}}$, the number of rows in the PE array ($\mathsf{PE}_{\mathsf{rows}}$), the number of columns in the PE array ($\mathsf{PE}_{\mathsf{col}}$), global buffer size ($\mathsf{GBS}$), the input feature map dimension ($\mathsf{A}$), the input channel count ($\mathsf{C}$), filter count ($\mathsf{F}$), kernel size ($\mathsf{K}$), stride ($\mathsf{S}$), and padding ($\mathsf{P}$). For ResNets, we add two more binary features that indicate whether the layer contains regular skip connection $\mathsf{RS}\in\{0,1\}$ and dotted skip connection $\mathsf{DS}\in\{0,1\}$. Similar to power and area models, we build latency models specific to each processing element type to capture the correlation between PE implementations to latency. We use these latency models to obtain the performance results by taking the inverse of latency estimations. Therefore, we refer to the inverse of latency as performance throughout the paper.

Figure~\ref{fig:model_selection} depicts the model selection methodology used in our framework. We compare the mean absolute percentage error (MAPE) and root mean square percentage error (RMSPE) jointly to properly tune the model parameters of our polynomial model which we also apply model selection techniques based on \textit{k}-fold cross validation \cite{kfold} to tune the degree of the polynomial. As it can be seen, both MAPE and RMSPE results continuously decrease with a two degree of the polynomial model until a five degree of the polynomial model. Then, both MAPE and RMSPE results increase and get significantly higher. Therefore, for the power, performance, and area models, we use five degree polynomial models as both MAPE and RMSPE results are negligibly small as shown in Figure~\ref{fig:model_selection}.

\section{Results}\label{sec:quidam_results}
In this section, we present power, performance, and area modeling results for each processing element type and perform a design space exploration on various DNN models such as VGG-16 \cite{vgg16}, ResNet-20, ResNet-34, ResNet-50, and ResNet-56 \cite{resnet50} on CIFAR-10, CIFAR-100, and ImageNet datasets to iterate through our framework to demonstrate the flexibility of \textit{QUIDAM} for future studies. 

\subsection{Power, Performance, and Area Model Accuracy}\label{sec:quidam_model_accuracy}

The power, performance, and area models detailed in Section \ref{sec:quidam_methodology} significantly speed up the design space exploration by 3-4 orders of magnitude. Indeed, \textit{QUIDAM} enables fast design exploration since it reduces the characterization process from days for synthesizing the RTL implementation and determining power, performance, and area, results to seconds for using the trained models.

Figure~\ref{fig:power_model}-\ref{fig:area_model} show the actual and estimated power, performance, and area results for each processing element type such as FP32, INT16, LightPE-1, and LightPE-2. Each data point in Figure~\ref{fig:power_model}-\ref{fig:area_model} corresponds to a different hardware accelerator configuration in the comprehensive design space. As shown by the results, \textit{QUIDAM}'s PPA models achieve high correlation to the actual PPA values. 

Figure~\ref{fig:power_model},\ref{fig:area_model} also show that the FP32 implementation has the highest area and power cost whereas LightPEs have the lowest area and power results when one processing element is considered. This shows the hardware-efficiency of LightPEs when compared to conventional PE implementations.

We also note that Figure~\ref{fig:power_model} and Figure~\ref{fig:area_model} show a higher correlation to actual power and area results than the corresponding chart for latency shown in Figure~\ref{fig:performance_model}. This is expected because the performance results depend on both hardware accelerator and deep neural network configurations whereas the power and the area models have only hardware accelerator features. Therefore, building a close to perfect performance model is more difficult given the feature space dimensionality and richness.

\subsection{Accelerator Design Space Exploration Results}\label{sec:quidam_dse_results}

\begin{figure*}[t]
  \centering
 \includegraphics[width=0.49\textwidth]{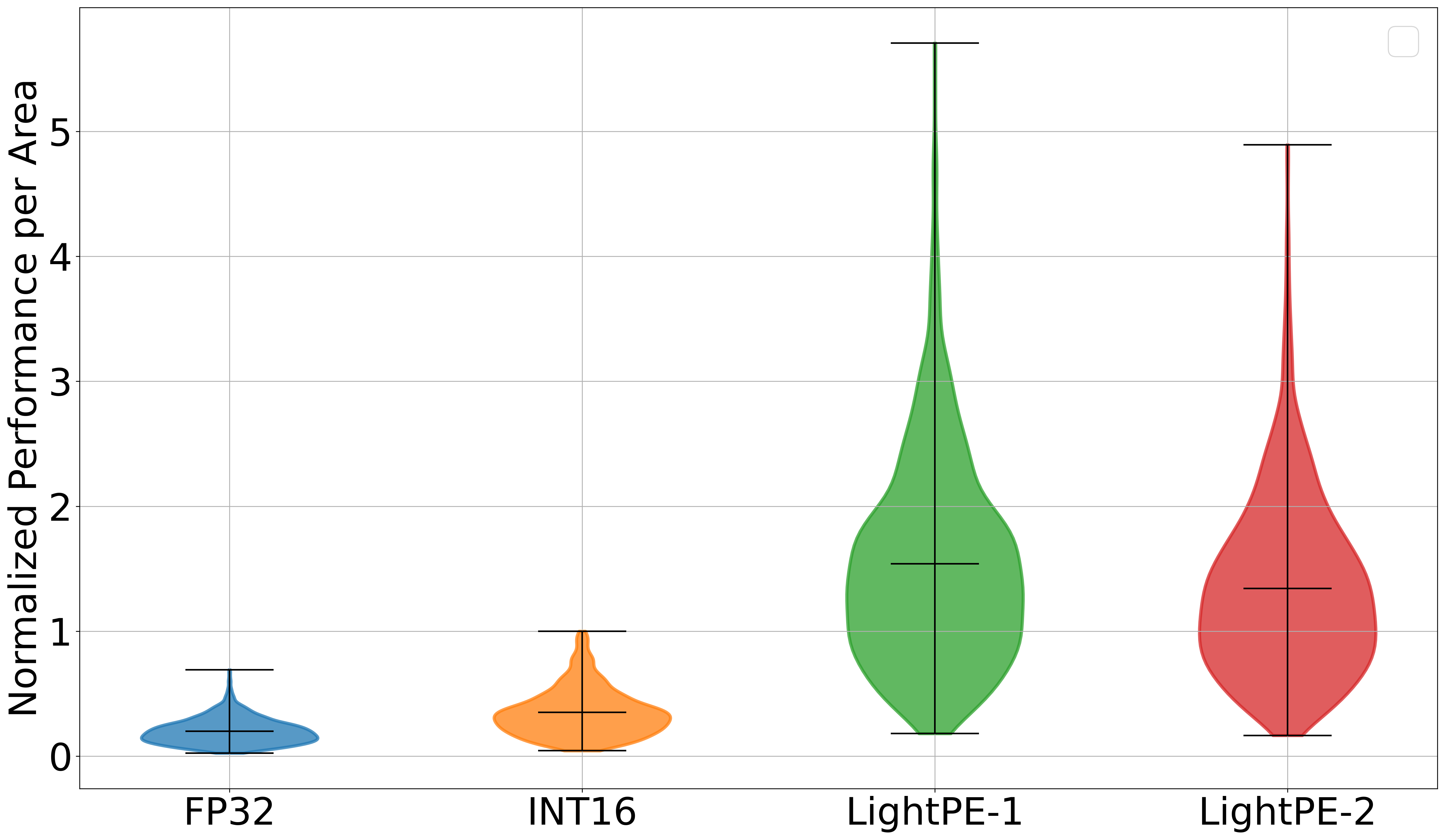}
  \includegraphics[width=0.49\textwidth]{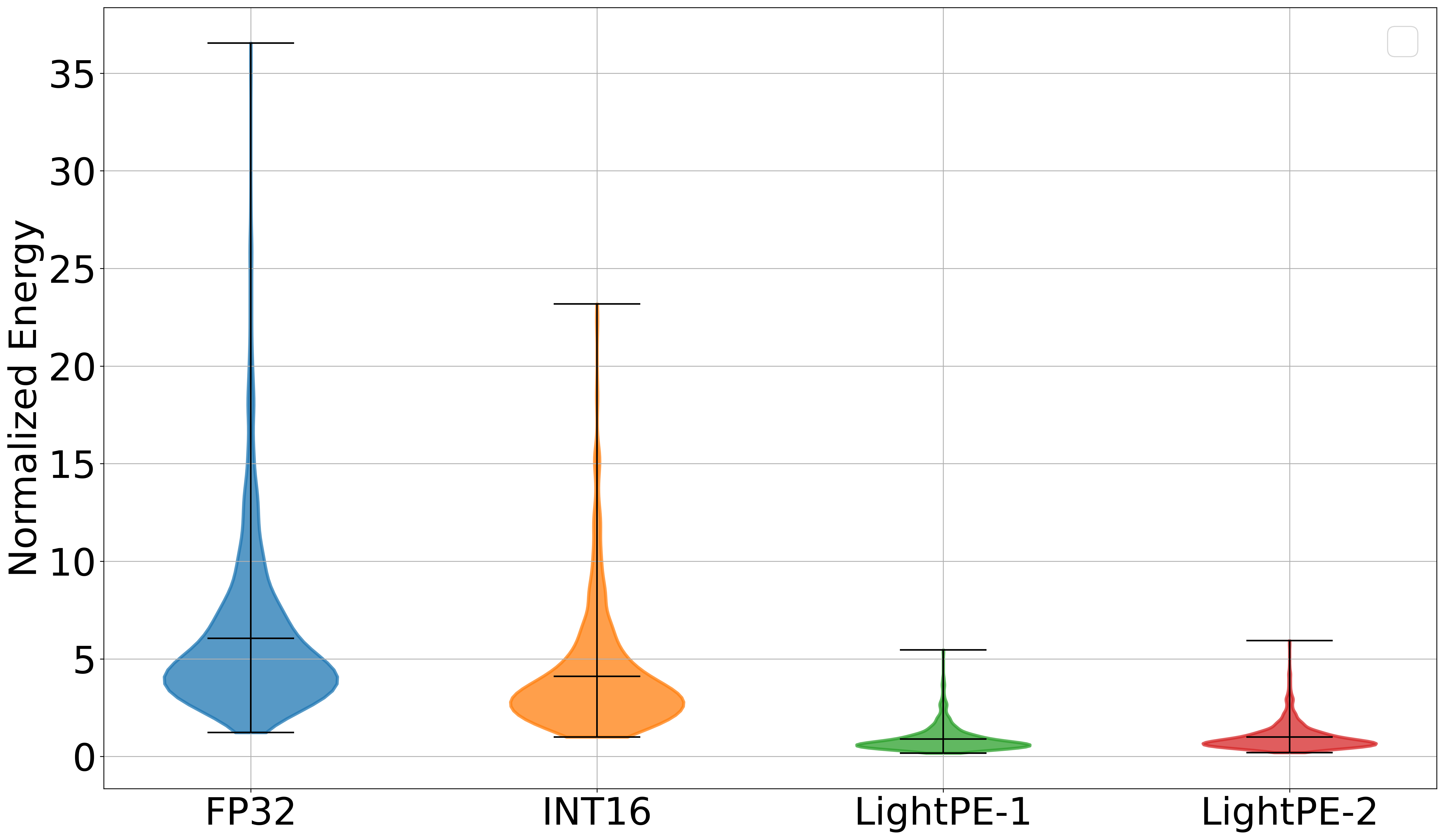}
 \caption{Violin plots showing the full distribution of normalized performance per area (left chart) and normalized energy (right chart) results with respect to the best INT16 configuration for various processing element types such as FP32, INT16, LightPE-1, and LightPE-2. Each plot shows the full spectrum results for each PE type and black bars show the minimum, the maximum, and the median values. LightPEs provide higher performance per area and energy-efficient designs when compared to FP32 and INT16 based designs.}
 \label{fig:violin}
\end{figure*}

To show the efficacy of LightPEs to conventional PE designs, we perform design space exploration on various DNN models such as VGG-16 \cite{vgg16}, ResNet-20, ResNet-34, ResNet-50, and ResNet-56 \cite{resnet50} on CIFAR-10, CIFAR-100, and ImageNet datasets as shown in Figure~\ref{fig:violin}.
We show the normalized performance per area and normalized energy results for each PE type with respect to the baseline INT16 based implementation with the highest performance per area for the given design space. Performance per area is a useful metric as different processing element implementations use different bit precision and this affects the required storage of these implementations. Therefore, we use performance per area as a comparison metric in our analysis. 

As it can be seen, LightPE implementations consistently outperform conventional INT16 and FP32 in both performance per area and energy objectives, which proves their efficacy in terms of hardware-efficiency. Figure~\ref{fig:violin} shows the full distribution including the minimum, the maximum, and the median results for each PE type. As it can be seen, LightPEs consistently provide higher performance per area and lower energy consuming design points as opposed to FP32 and INT16 based designs. Specifically, LightPE-1 and LightPE-2 achieve $4.8 \times$ and $4.1 \times$ more performance per area and $4.7 \times$ and $4 \times$ less energy on average across VGG-16, ResNet-20, and ResNet-56 workloads on CIFAR-10/CIFAR-100 and VGG-16, ResNet-34, and ResNet-50 workloads on ImageNet datasets when compared to the best INT16 hardware configuration, respectively. On the other hand, INT16 baseline implementation achieves  $1.8 \times$ more performance per area and $1.5 \times$ less energy on average when compared to the best FP32 configuration.

These conclusions hold for all the models and the datasets considered in this work such as VGG-16, ResNet-20, ResNet-34, ResNet-50, and ResNet-56 thereby showing that the benefits of using lower precision generalize across a variety of models. 
We conclude that different bit precisions and PE types can lead to significantly different performance per area and energy results which are two critical metrics for machine learning and systems community strives to improve upon.

\subsection{Pareto-Optimality for Accuracy and Performance per Area}\label{sec:quidam_pareto_perfarea}
% \textbf{Pareto-Optimality for Accuracy and Performance per Area}

\begin{figure*}[t]
  \centering
 \includegraphics[width=0.49\textwidth]{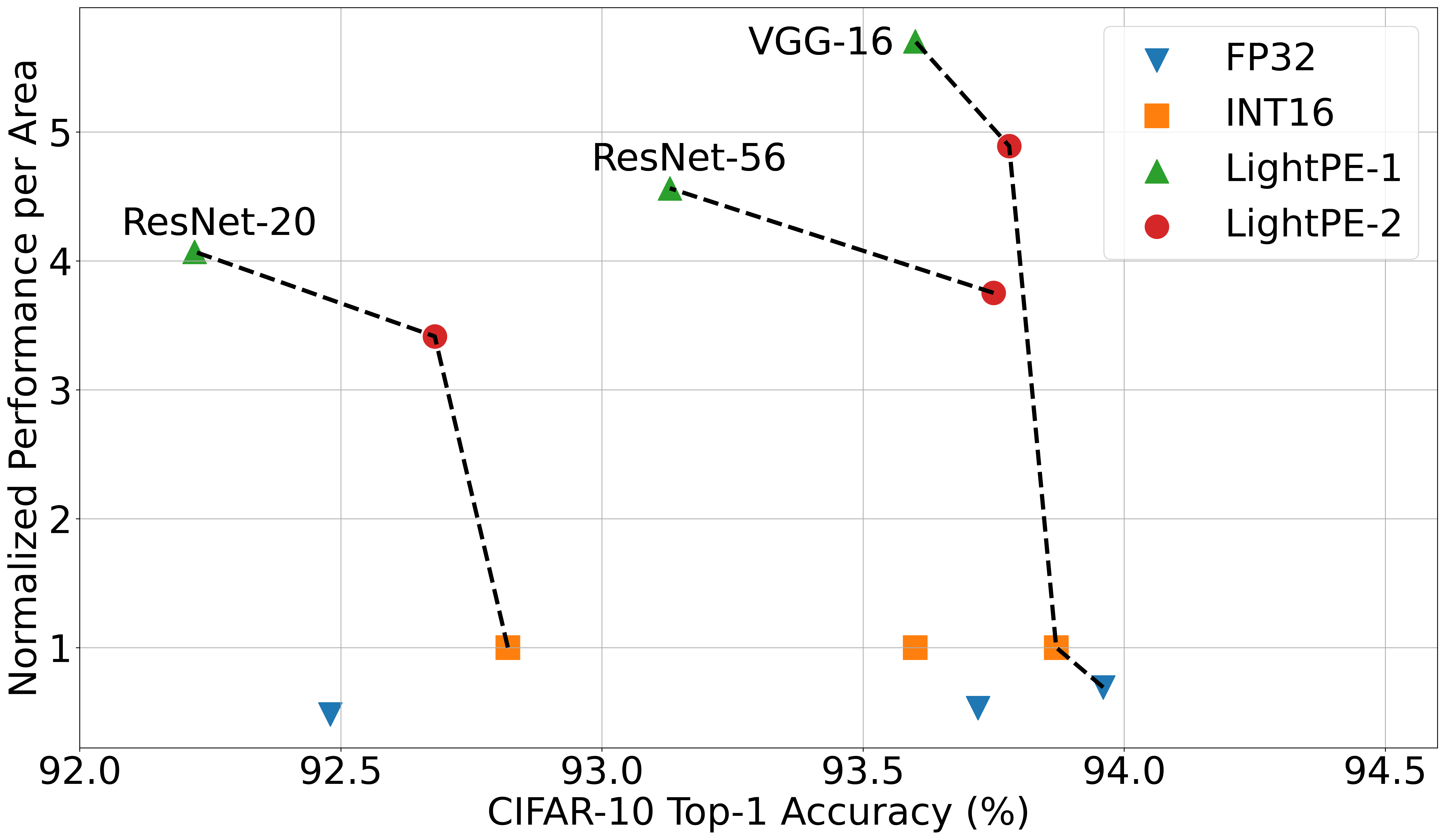}
  \includegraphics[width=0.49\textwidth]{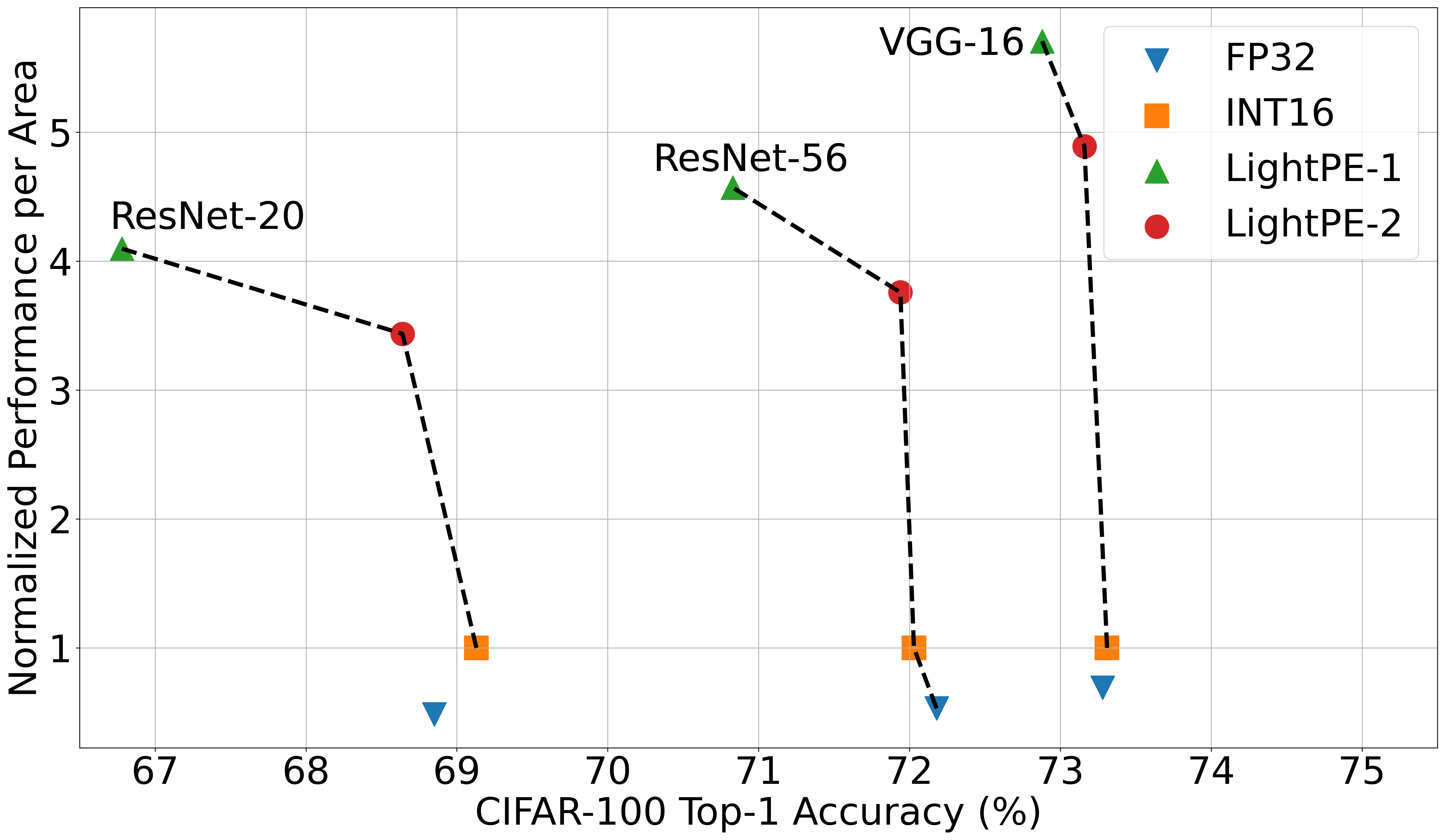}
 \caption{Normalized performance per area and top-1 accuracy results for various processing element types such as FP32, INT16, LightPE-1, and LightPE-2 for CIFAR-10 (left chart) and CIFAR-100 (right chart). Each data point corresponds to the hardware configuration with the highest performance per area for the corresponding processing element type. Pareto-front is shown with a dashed line for each DNN model. LightPEs are consistently on Pareto-front for various DNN models.}
 \label{fig:pareto_perf_area}
\end{figure*}

To show the accuracy and performance per area trade-off for different processing element types, we perform a Pareto-front analysis by training VGG-16, ResNet-20, and ResNet-56 models for CIFAR-10 and CIFAR-100 datasets. For both datasets, we perform five runs for each DNN model and processing element type and plot the mean top-1 accuracy results. The training recipe for both CIFAR-10/CIFAR-100 datasets follows prior art \cite{yang2018soft,Chin2020CVPR} which uses stochastic gradient descent with nesterov momentum, weight decay 0.0005, batch size 128, 0.1 initial learning rate with decrease by $5 \times$ at epochs 60, 120, and 160, and train for 200 epochs in total. We note that this training recipe is tuned for full-precision models. Therefore, the accuracy results for LightPE variants might be higher with proper hyperparameter tuning.

\begin{figure*}[t]
  \centering
 \includegraphics[width=0.49\textwidth]{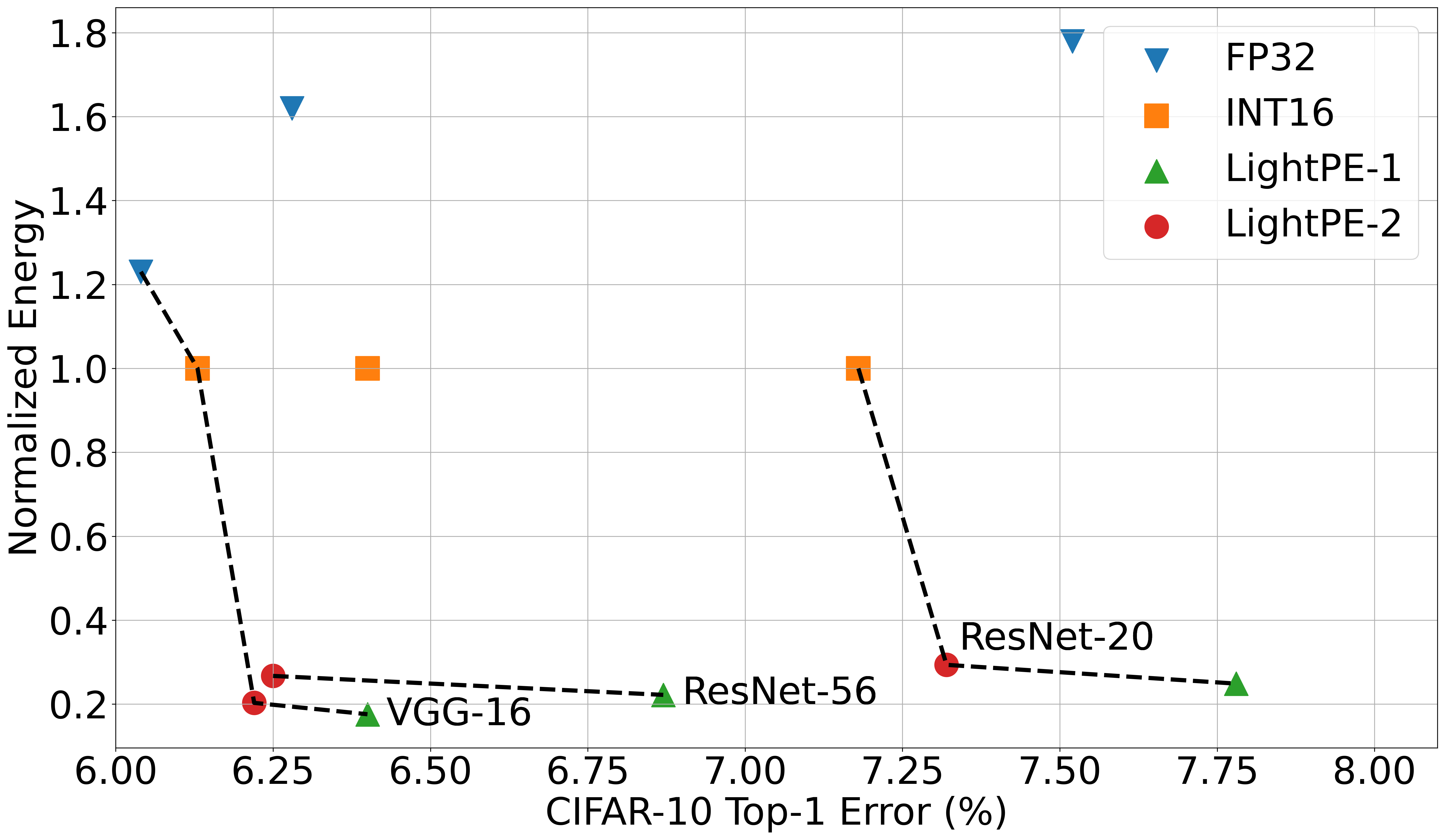}
  \includegraphics[width=0.49\textwidth]{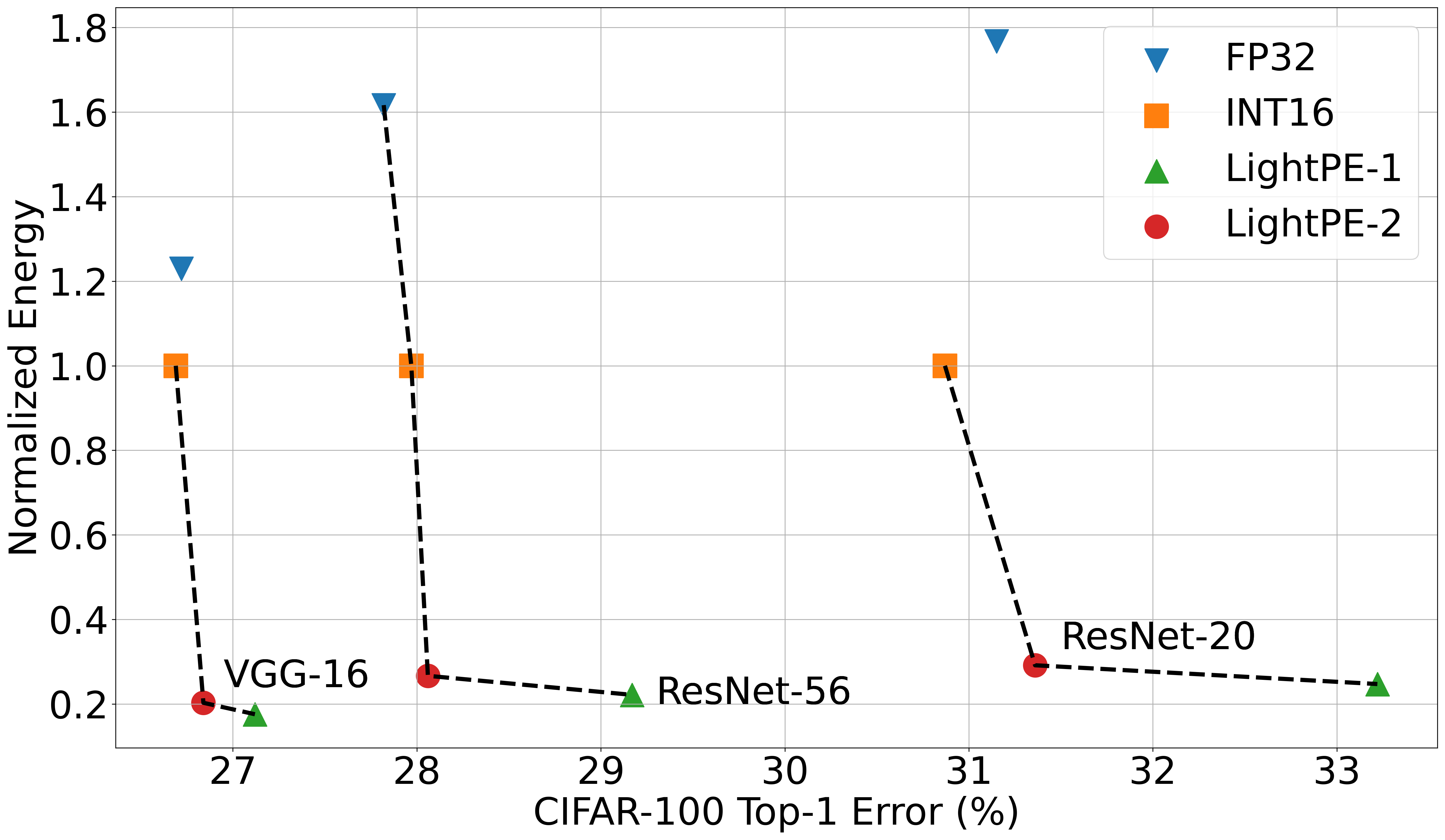}
 \caption{Normalized energy and top-1 error results for various processing element types such as FP32, INT16, LightPE-1, and LightPE-2 for CIFAR-10 (left chart) and CIFAR-100 (right chart). Each data point corresponds to the hardware configuration with the lowest energy for the corresponding processing element type. Pareto-front is shown with a dashed line for each DNN model. LightPEs are consistently on Pareto-front for various DNN models.}
%  \vspace{-3mm}
 \label{fig:pareto_energy}
\end{figure*}

Figure~\ref{fig:pareto_perf_area} shows the normalized performance per area and accuracy results for FP32, INT16, LightPE-1, and LightPE-2. Performance per area results are normalized with respect to the best INT16 configuration for each DNN model. We plot the hardware configurations with the highest performance per area results for each processing element type. Next, we perform a Pareto-front analysis among different processing element types and show the Pareto-frontier with a dashed line for each DNN model. As it can be seen, LightPEs are consistently on the Pareto-front for various DNN models and datasets, whereas FP32 and INT16 based designs are occasionally dominated by LightPE variants mostly due to LightPE implementations pushing the Pareto-frontier by being more hardware-efficient in terms of performance per area and energy than FP32 and INT16 based designs. Moreover, in certain situations such as CIFAR-10 accuracy results LightPE-2 based design dominates FP32 and both INT16 and FP32 not only in hardware-efficiency metrics but also in accuracy results for ResNet-20 and ResNet-56 models, respectively.
To sum up, LightPE-1 and LightPE-2 achieve on par accuracy results with FP32 and INT16 while achieving up to $5.7 \times$ and $4.9 \times$ more performance per area when compared to INT16 configuration, respectively.

\subsection{Pareto-Optimality for Accuracy and Energy}\label{sec:quidam_pareto_energy}
% \paragraph{Pareto-Optimality for Accuracy and Energy}

We also perform a Pareto-front analysis for accuracy and energy results. We follow the same training methodology explained in Section 4-2. Figure~\ref{fig:pareto_energy} shows the normalized energy and accuracy results for FP32, INT16, LightPE-1, and LightPE-2 based designs. Energy results are normalized with respect to the best INT16 configuration for each DNN model. As it can be seen, LightPEs are systematically on Pareto-front for various DNN models and datasets. Specifically, LightPE-1 and LightPE-2 achieve $4.7 \times$ and $4 \times$ less energy on average across different workloads and datasets when compared to INT16 configuration, respectively. In addition, we note that as the model complexity increases, the accuracy gap between LightPEs and conventional FP32 and INT16 based designs decreases. Thus, we conclude that our proposed LightPEs have promising results for training larger models with negligible accuracy loss while achieving significant performance per area and energy improvements.

\begin{table*}[t]
\centering
\caption{{{{Pareto-Optimal Results}}}}
\label{table:cifar10_summary}
\resizebox{0.9\textwidth}{!}{%
\begin{tabular}{||c|c|cc|c|c||}
\hline
\multirow{2}{*}{Network} & \multirow{2}{*}{PE Type} & \multicolumn{2}{c|}{Accuracy (\%)} & \multirow{2}{*}{Energy} & \multirow{2}{*}{Performance per Area} \\
& & CIFAR-10 & CIFAR-100 & &\\
\hline\hline
\multirow{4}{*}{VGG-16} & FP32 & \textbf{93.96} & 73.28 & $1.2 \times$ & $0.69 \times$ \\ \cline{2-6} 
 & INT16 & 93.87 & \textbf{73.31} & $1 \times$  & $1 \times$ \\ \cline{2-6} 
 & LightPE-2 & 93.78 & 73.16 & $0.20 \times$ & $4.9 \times$ \\ \cline{2-6}
 & LightPE-1 & 93.60 & 72.88 & $\textbf{0.18} \times$ & $\textbf{5.7} \times$ \\ \hline\hline 
\multirow{4}{*}{ResNet-20} & FP32 & 92.48 & 68.85 & $1.8 \times$ & $0.48 \times$ \\ \cline{2-6} 
 & INT16 & \textbf{92.82} & \textbf{69.13} & $1 \times$ & $1 \times$ \\ \cline{2-6} 
 & LightPE-2 & 92.68 & 68.64 & $0.29 \times$ & $3.4 \times$ \\ \cline{2-6} 
 & LightPE-1 & 92.22 & 66.78 & $\textbf{0.25} \times$ & $\textbf{4.1} \times$ \\ \hline\hline
\multirow{4}{*}{ResNet-56} & FP32 & 93.72 & \textbf{72.18} & $1.6 \times$ & $0.53 \times$ \\ \cline{2-6} 
 & INT16 & 93.60 & 72.03 & $1 \times$ & $1 \times$ \\ \cline{2-6} 
 & LightPE-2 & \textbf{93.75} & 71.94 & $0.27 \times$ & $3.8 \times$ \\ \cline{2-6} 
 & LightPE-1 & 93.13 & 70.83 & $\textbf{0.22} \times$ & $\textbf{4.6} \times$ \\ \hline
\end{tabular}%
}
\end{table*}

\begin{table}[t]
\centering
\caption{{{Clock frequency values of \textit{QUIDAM} generated designs with different PE types}}}
\label{table:frequency}
\resizebox{0.35\textwidth}{!}{%
\begin{tabular}{||c|c||}
\hline
PE Type & Clock Frequency \\
\hline \hline
FP32 & 275 MHz \\
INT16 & 285 MHz \\
LightPE-2 & 435 MHz \\
LightPE-1 & 455 MHz \\
\hline
\end{tabular}%
}
\end{table}

We summarize our findings in Table~\ref{table:cifar10_summary} which shows the Pareto-optimal results for each PE type for model accuracy and hardware-efficiency metrics such as performance per area and energy for VGG-16, ResNet-20, and ResNet-56 models on CIFAR-10 and CIFAR-100 datasets. Our results show that LightPE implementations provide on par accuracy results across models and datasets and improve the hardware-efficiency in terms of energy and performance per area when compared to FP32 and INT16 based designs. 
As it can be seen from Table~\ref{table:cifar10_summary}, LightPEs consistently dominate FP32 and INT16 based designs in terms of hardware-efficiency metrics such as energy and performance per area. In addition, although we only claim that LightPEs can achieve similar accuracy to FP32 and INT16 designs, in certain situations such as ResNet-20 and ResNet-56 for CIFAR-10 dataset, the LightPE-2 based design achieves higher accuracy than the FP32 based design for ResNet-20, and both FP32 and INT16 based designs for ResNet-56, respectively.

Furthermore, Table \ref{table:frequency} shows the clock frequency values found by \textit{QUIDAM} for designs with different PE types. We note that LightPE based implementations provide up to $1.7 \times$ and $1.6 \times$ speedup when compared to FP32 and INT16 based designs, respectively. 
In addition, we note that LightPE-2 and LightPE-1 implementations achieve 435 MHz and 455 MHz in 45 nm technology node \cite{freepdk}, respectively. As the Eyeriss design reports its core clock frequency as 200 MHz and it utilizes 65 nm technology node \cite{eyeriss_ssc}, we apply the prominent technology scaling rules to make a fair comparison among different designs. Based on these scaled calculations \cite{tech_scaling}, we note that \textit{QUIDAM} finds LightPE implementations that are $1.5 \times$ to $1.6 \times$ faster when compared to Eyeriss \cite{eyeriss_ssc} design. Moreover, with the same INT16 based implementation, the \textit{QUIDAM} generated DNN accelerator configuration achieves similar clock frequency (197 MHz) after technology scaling.

\subsection{DNN Accelerator and Model Co-Exploration}\label{sec:quidam_coexploration}

\begin{table}[t]
\centering
\caption{{{Search space for neural architectures used in DNN accelerator and model co-exploration analysis}}}
\label{table:nas-space}
\resizebox{0.65\textwidth}{!}{%
\begin{tabular}{||c|c|c||}
\hline
Block & Number of Repetitions & Channels\\
\hline \hline
Conv-BN-ReLU & \{1,2\} & \{40, 48, 56, 64\}\\
MaxPool & 1 & N/A\\
Conv-BN-ReLU & \{1,2\} & \{80, 96, 112, 128\}\\
MaxPool & 1 & N/A\\
Conv-BN-ReLU & \{1,2,3\} & \{160, 192, 224, 256\}\\
MaxPool & 1 & N/A\\
Conv-BN-ReLU & \{1,2,3\} & \{320, 384, 448, 512\}\\
MaxPool & 1 & N/A\\
Conv-BN-ReLU & \{1,2,3\} & \{320, 384, 448, 512\}\\
MaxPool & 1 & N/A\\
\hline
\end{tabular}%
}
\end{table}

So far, we have demonstrated the usefulness of our proposed framework by mainly varying the hardware architecture given some commonly adopted neural network designs. In this subsection, we demonstrate the generalizability of the proposed framework by co-exploring the design space of both hardware configurations and neural network architectures. To do so, we need an accuracy model for neural architectures in addition to the proposed power, performance, and area hardware cost models to rapidly iterate over different DNN models and perform a DNN accelerator and model co-exploration analysis. We note that an accuracy proxy model is needed since the search space of DNN architectures is extremely large (hundreds of thousands) which would require an untenable cost of training. To this end, we adopt the weight-sharing evaluation technique to estimate the accuracy of a candidate neural network architecture~\cite{li2020random,guo2020single}. More specifically, we first define a neural architecture search space for an existing VGG-16 network as shown in Table~\ref{table:nas-space}. The neural architecture search space used in our analysis is composed of Conv-BN-ReLU and MaxPool blocks with number of repetitions of each block ranging from 1 to 3, and number of channels ranging from 40 to 512 based on the layer number. It can be observed that the largest configuration is a VGG-16 architecture and there are smaller variants to be searched. The baseline VGG-16 model can be achieved by choosing the largest number of repetitions per block and the largest number of channels available in the search space shown in Table~\ref{table:nas-space}. The entire search space contains 110,592 candidate neural network architectures which is found by multiplying the number of all possible choices in each step of candidate architecture search. To obtain an accuracy predictor, we train the neural network by randomly sampling an architecture from the search space for each batch while sharing the weights with the largest neural network architecture~\cite{li2020random,guo2020single}. After the training is done, we randomly select $1,000$ network architectures and directly evaluate their accuracy on the validation set as our output for the accuracy predictor.

\begin{figure*}[t]
  \centering
 \includegraphics[width=0.49\textwidth]{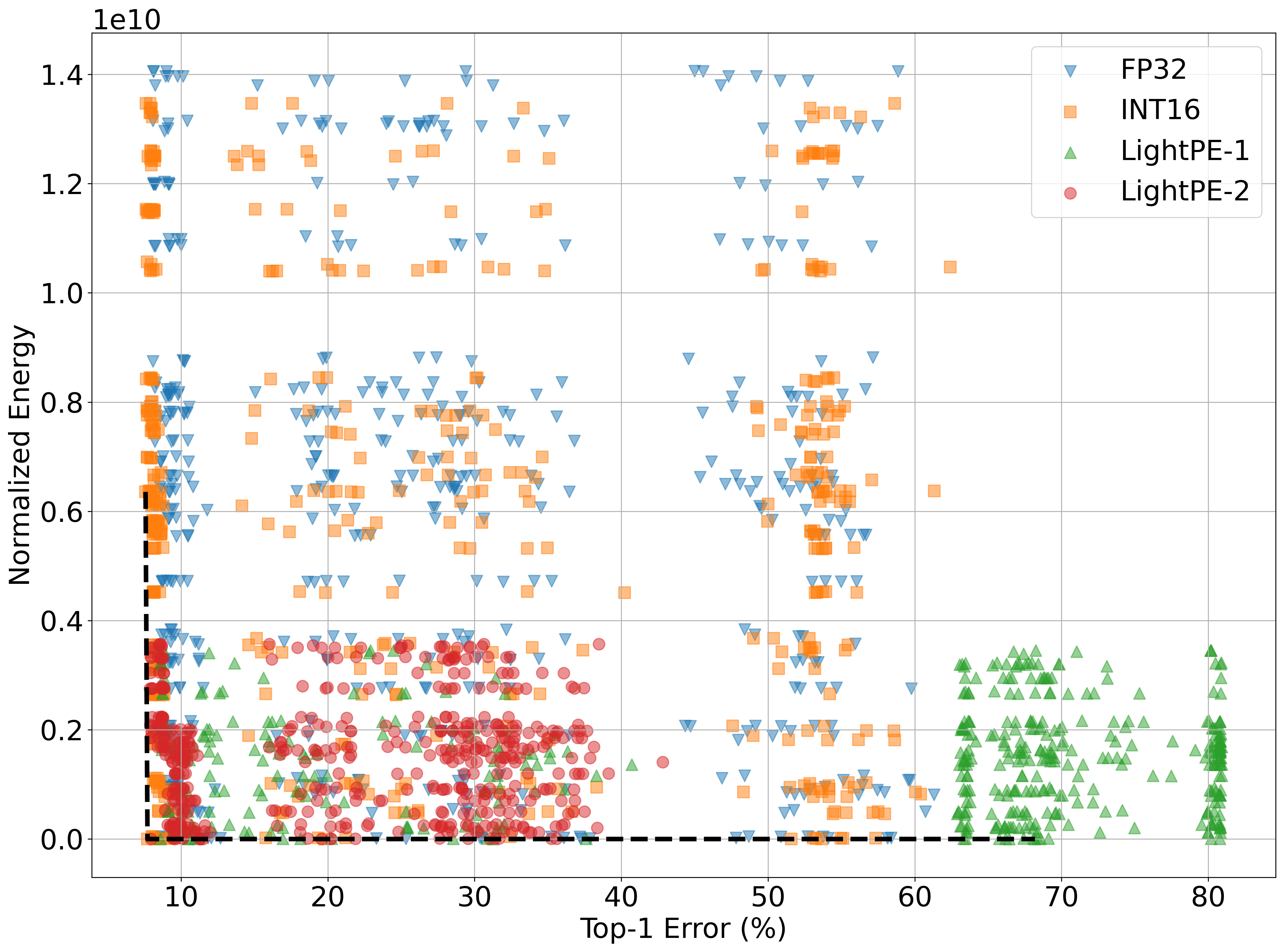} %norm_power_pareto
  \includegraphics[width=0.49\textwidth]{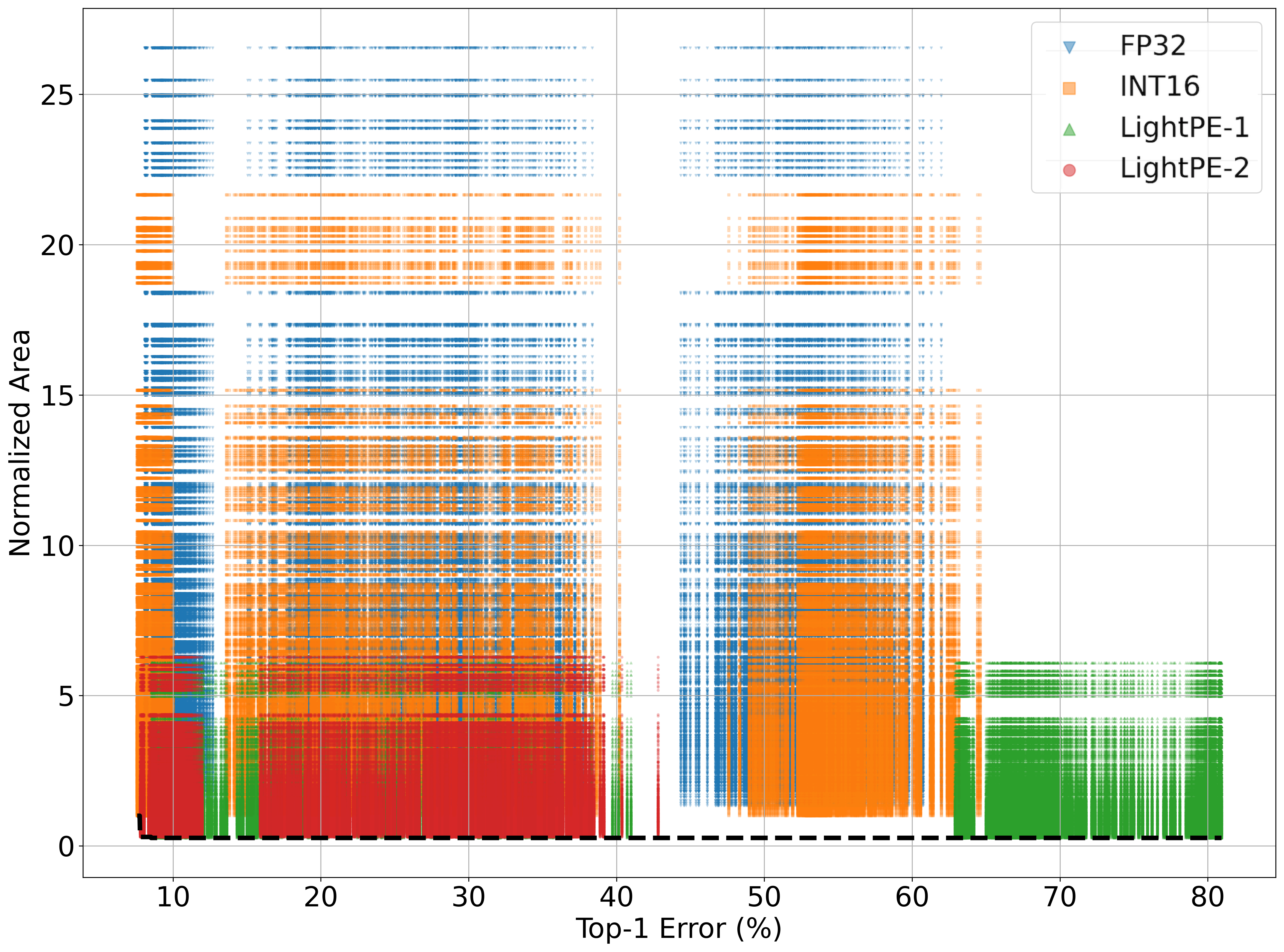}
 \caption{Normalized energy (left chart) and normalized area (right chart) vs. top-1 error results for various DNN configurations and processing element types such as FP32, INT16, LightPE-1, and LightPE-2. Each data point corresponds to a different hardware and DNN architecture pair which are normalized to the minimum energy (left chart) and the minimum area (right chart) pair in the INT16 design space. Pareto-front for the co-exploration space is shown with a dashed line. As it can be seen, LightPEs are consistently on Pareto-front even when DNN accelerator and model configurations are co-explored.}
 % \caption{Normalized power (left chart) and normalized area (right chart) vs. top-1 error results for various DNN configurations and processing element types such as FP32, INT16, LightPE-1, and LightPE-2. Each data point corresponds to a different hardware and DNN architecture pair which are normalized to the maximum power (left chart) and the minimum area (right chart) pair in the INT16 design space. Pareto-front for the co-exploration space is shown with a dashed line. As it can be seen, LightPEs are consistently on Pareto-front even when DNN accelerator and model configurations are co-explored.}
%   \vspace{-2mm}
 \label{fig:power_area_pareto}
\end{figure*}

Figure~\ref{fig:power_area_pareto} shows the normalized energy and the normalized area \textit{vs.} top-1 model error results for various DNN accelerator and model pairs on the CIFAR-10 dataset. We randomly sample accelerator configurations and use $1000$ DNN models. We then evaluate each accelerator and model pairs in terms of Pareto-optimality. Performance per area and energy results are normalized with respect to the minimum energy and area point in the INT16 design space, respectively. Figure~\ref{fig:power_area_pareto} shows that LightPEs are consistently on the Pareto-front even when the DNN accelerator and model configurations are co-explored which shows the efficacy of LightPEs not only in a few commonly adopted DNN models but in a generalized DNN accelerator and model co-design space.

Based on these analyses and results, we conjecture that \textit{QUIDAM} can successfully provide a wide range of DNN accelerator and model pairs based on different needs in terms of accuracy and critical hardware-efficiency metrics such as performance per area and energy. Therefore, we conclude that \textit{QUIDAM} can be used for DNN accelerator and model co-design as it incorporates quantization-aware hardware and PPA models which significantly speeds up the co-design efforts.

\section{Conclusion}\label{sec:quidam_conclusion}
In this work, we present \textit{QUIDAM}, a quantization-aware highly parameterized DNN accelerator and model co-exploration framework. Our framework can foster future research on design space co-exploration of DNN accelerators for various design choices such as bit precision, processing element type, scratchpad size of processing elements, global buffer size, device bandwidth, number of total processing elements in the the design, and DNN configurations. 
Our results show that different bit precisions and processing element types lead to significant differences in terms of performance per area and energy. Specifically, LightPE-1 and LightPE-2 achieve $4.8 \times$ and $4.1 \times$ more performance per area and $4.7\times$ and $4 \times$ energy improvement on average when compared to the best INT16 hardware configuration, respectively. We also show that our proposed LightPEs consistently achieve Pareto-optimal results in terms of accuracy and performance per area and energy for commonly used DNN models as well as when DNN accelerator and model configurations are co-explored. Therefore, design space co-exploration of quantization-aware DNN accelerators and models merits a meticulous analysis that takes these factors into account. 

% Identification of funding sources and other support, and thanks to
% individuals and groups that assisted in the research and the
% preparation of the work should be included in an acknowledgment
% section, which is placed just before the reference section in your
% document.

% This section has a special environment:
% \begin{verbatim}
%   \begin{acks}
%   ...
%   \end{acks}
% \end{verbatim}
% so that the information contained therein can be more easily collected
% during the article metadata extraction phase, and to ensure
% consistency in the spelling of the section heading.

% Authors should not prepare this section as a numbered or unnumbered {\verb|\section|}; please use the ``{\verb|acks|}'' environment.

\begin{acks}
This research was supported in part by NSF CCF Grant No. 1815899 and NSF CSR Grant No. 1815780.
\end{acks}

\bibliographystyle{ACM-Reference-Format}
\bibliography{refs}

% \appendix

\end{document}